\documentclass[sigconf]{acmart}

\usepackage{caption}
\usepackage{algorithm}
\usepackage{algorithmic}
\usepackage{subcaption}
\usepackage{amsmath}
\usepackage{graphicx}
\usepackage{enumitem}

\AtBeginDocument{%
  }

\copyrightyear{2023}
\acmYear{2023}
\setcopyright{rightsretained}
\acmConference[AIES '23]{AAAI/ACM Conference on AI, Ethics, and Society}{August 8--10, 2023}{Montréal, QC, Canada}
\acmBooktitle{AAAI/ACM Conference on AI, Ethics, and Society (AIES '23), August 8--10, 2023, Montréal, QC, Canada}
\acmDOI{10.1145/3600211.3604705}
\acmISBN{979-8-4007-0231-0/23/08}

\begin{document}

\title{Factoring the Matrix of Domination: A Critical Review and Reimagination of Intersectionality in AI Fairness}

\author{Anaelia Ovalle}
\affiliation{%
  \institution{Department of Computer Science}
  \city{University of California, Los Angeles}
  \country{}
}

\author{Arjun Subramonian}
\affiliation{%
  \institution{Department of Computer Science}
  \city{University of California, Los Angeles}
  \country{}
}

\author{Vagrant Gautam}
\affiliation{%
  \institution{Spoken Language Systems}
  \city {Saarland University}
  \country{}
  }

\author{Gilbert Gee}
\affiliation{%
  \institution{Department of Community Health}
  \city {University of California, Los Angeles}
  \country{}
  }

\author{Kai-Wei Chang}
\affiliation{%
  \institution{Department of Computer Science}
  \city {University of California, Los Angeles}
  \country{}
  }
\renewcommand{\shortauthors}{Ovalle et al.}

\begin{abstract}
Intersectionality is a critical framework that, through inquiry and praxis, allows us to examine how social inequalities persist through domains of structure and discipline. Given AI fairness' raison d'être of ``fairness,'' we argue that adopting intersectionality as an analytical framework is pivotal to effectively operationalizing fairness. Through a critical review of how intersectionality is discussed in 30 papers from the AI fairness literature, we deductively and inductively: 1) map how intersectionality tenets operate within the AI fairness paradigm and 2) uncover gaps between the conceptualization and operationalization of intersectionality. We find that researchers overwhelmingly reduce intersectionality to optimizing for fairness metrics over demographic subgroups. They also fail to discuss their social context and when mentioning power, they mostly situate it only within the AI pipeline. We: 3) outline and assess the implications of these gaps for critical inquiry and praxis, and 4) provide actionable recommendations for AI fairness researchers to engage with intersectionality in their work by grounding it in AI epistemology.
\end{abstract}

\ccsdesc[500]{Social and professional topics~Computing / technology policy}
\ccsdesc[300]{Computing methodologies~Artificial intelligence}

\keywords{fairness, intersectionality, artificial intelligence, literature review}

\maketitle

\section{Introduction}
\label{sec:intro}
Artificial intelligence (AI) fairness research is critical to the development of just AI. Work in this space consistently urges researchers and engineers alike to consider notions of fairness defined over model predictions. These notions vary across conceptualization (e.g., group, individual fairness \cite{Binns2019OnTA}) and operationalization (e.g., pre/in/post-processing \cite{barocas-hardt-narayanan}) \cite{Jacobs2019MeasurementAF}; nevertheless, the literature generally agrees on the goal of minimizing negative outcomes across demographic groups, including groups associated with multiple, ``intersectional'' demographic attributes (e.g., Black women) \cite{wang2022TI}. However, \citet{P3} observes that AI fairness papers often narrowly interpret intersectional subgroup fairness as intersectionality, the critical framework from which the term originates \cite{collins2020intersectionality, kong2022intersectionally}. This myopic conceptualization of intersectionality has non-trivial consequences for just AI design and epistemology (i.e., ways of knowing).

The term \textit{intersectionality} describes a traveling framework of critical inquiry and praxis (i.e., practical action beyond mere academic theorizing) intended to examine interlocking mechanisms of structural oppression (e.g., racist policy \cite{Kendi_2023}) which produce inequality \cite{collins2020intersectionality}. Critical inquiry into the formation of inequalities generates knowledge that can inform strategies for combating them, which is often referred to as praxis. Generating knowledge that illuminates the underlying mechanisms of oppressive systems is a shared objective among critical disciplines, such as feminist, antiracist, and decolonial studies, and is rooted in a history of resistance \cite{collins2019intersectionality}.
Critical disciplines thus do not decouple reclaiming knowledge from reclaiming power. This is in contrast to disciplines rooted in colonial epistemology, e.g., science.
Upon initial examination, science offers universal, empirically-grounded explanations for natural phenomena; however, science is rooted in colonialism through its imposing of a ``a positivist paradigm\footnote{Knowledge as a result of ``neutral'' and quantifiable observation. This paradigm strictly relies on only measurement and reason \cite{nottinghamUnderstandingPragmatic}.}
approach to research on the colonies and other oppressed groups'' \cite{chilisa2019indigenous}. According to scientific colonialism, the researcher has ``unlimited rights of access to source[s] of information belonging to [a] population,'' where data collection and knowledge formation reflects \textit{the one reality} the researcher understands \cite{cameron2004evidence, cram2004kaupapa}. Indigenous knowledge is erased as dominant knowledge systems are imposed, preventing Indigenous people from creating and sharing their own knowledge and perspectives. Consequently, disciplines rooted in colonial epistemology often assimilate prevailing knowledge systems that perpetuate the erasure of knowledge \cite{chilisa2019indigenous, fanon2004wretched, cowan1972francis}.

The epistemologies of AI research are not divorced from scientific colonialism's legacy. Intersectionality may be used to critically examine AI research methodologies, so that ``the world-views of those who have suffered a long history of oppression and marginalization are given space to communicate from their frames of reference'' \cite{chilisa2019indigenous}.
Intersectionality promotes grappling with ``how individuals and groups who are subordinated within varying systems of power might survive and resist their oppression,'' thereby empowering communities to criticize the injustices they experience \cite{collins2019intersectionality}. In the face of epistemic violence (e.g., the erasure of Indigenous knowledge), intersectionality erects a new form of epistemic resistance: knowledge production. Frameworks to articulate social inequalities have been integral to the survival of communities at the margins. Similarly, intersectionality, by enabling researchers to observe and articulate disparities, may break the epistemic molds ``researchers are placed in so they may operate differently'' \cite{chilisa2019indigenous}.

In the context of AI fairness, intersectionality is less about getting technology right (e.g., establishing fairness constraints for a model); it is more about interrogating the social reality which drives AI oppression, so we can then make technology better. Crenshaw uses the term intersectionality as a metaphor to speak on how ``different systems of oppression overlap,'' but more importantly emphasizes that neglecting the convergence of these structures would cause rhetorical and identity politics to abandon issues and people who are actually affected by these intersecting ``systems of subordination'' \cite{berger2010intersectional}. Intersectionality thereby challenges the sociopolitical amnesia which frames subgroup fairness as solely a technical problem \cite{wang2022TI}. We do not reject subgroup fairness outright; rather, we share this example to challenge the AI fairness community to expand its engagement with intersectionality.
To operationalize AI fairness with an intersectional lens, it is vital to first illuminate underexplored gaps between intersectionality and existing AI fairness literature. To this end, we ask: (1) how is intersectionality discussed in AI fairness literature?; (2) to what extent does this discussion change based on computer science (CS) methodology?; (3) where are the largest gaps in conceptualizing and operationalizing intersectionality for advancing social justice?; (4) what tensions exist in leveraging these gaps for just AI design?; and (5) what do these findings tell us about opportunities for more just AI? To answer these questions, we contribute the following:
\begin{enumerate}
    \item Identify a growing body of AI fairness papers related to intersectionality (\S\ref{sec:review methods}) and examine their conceptions of the critical framework in contrast to core intersectionality literature (\S\ref{sec:intersect-overview}).
    \item Create guiding questions to critically assess the use of intersectionality as a lens to operationalize AI fairness  (Table \ref{tbl:guiding_questions}).
    \item Use our findings to analyze where gaps remain in AI fairness papers' use of intersectionality, provide recommendations towards addressing these gaps, and comment on the structural forces that may contribute to these observed norms (\S\ref{sec: deductive}, \S\ref{sec: inductive}).
\end{enumerate}

The majority of the papers we review approach intersectionality from the narrow perspective of subgroup fairness. Through a deductive lens in \S\ref{sec: deductive}, we find that intersectionality engagement varies significantly depending on how it is situated within the AI pipeline, how sources of biases are described, and what CS research epistemologies are invoked. Inductively in \S\ref{sec: inductive}, we find that even when researchers center intersectionality literature, there is little engagement with the framework itself, evidenced by a lack of described social context, little discussion of power and relations between structures, questionable citational practices, and a disjointed sense of social justice praxis.

Our paper does not concern itself with claiming that intersectionality must take a particular form within AI fairness. Rather, we center intersectionality as an ``analytical sensibility'' \cite{Cho_Crenshaw_McCall_2013, collins2020intersectionality},
which when activated, can sharpen and transform the tools in the AI fairness researcher's toolbox. This, we argue, is key to justice-centric AI development. We further seek to dispel the misconception that social science disciplines have no place in STEM \cite{raji2021exclusionary, mohamed2020decolonial}. Educated in CS, we equip the AI fairness researcher of similar training who is committed to justice with concrete ways of using their training in AI to exercise critical praxis. In this way, we hope to disrupt deep-rooted indifferences to social reality, ``a powerful force that is perhaps more dangerous than malicious intent'' \cite{benjamin2019race}.

\textit{Positionality Statement }
\label{sec:positionality}
All but one author of this paper are formally trained primarily as computer scientists, with additional training in gender theory, criticial social theories, criminology, linguistics, and related fields. One author is a social scientist who confronts issues of social inequities in both everyday life and their scholarship, necessitating an intersectional and life course perspective. All authors have informal training in queer studies through activism and advocacy. As such, our backgrounds influence this work's design, decisions, and development. All authors are located in the US or Europe, but have diasporic links to other social contexts; we do our best to position our work in a global context. We write this to empower individuals across both academia and industry research to critically engage with AI fairness paradigms. Therefore, our recommendations are articulated in a way that can be operationalized, though they are transferrable to other audiences.
We position ourselves within a social justice ethos informed by decolonial theory, and that champions equity over equality as well as reparations to correct historical injustices.

\vspace*{-3pt}
\section{Related Works}
\label{sec:related-works}
We are not the first to champion or critique intersectional praxis in AI fairness, let alone more broadly. Several works across disciplines including psychology and CS have advocated the use of intersectionality frameworks or discussed the misappropriation thereof \citep[\textit{inter~alia}]{Cole2009IntersectionalityAR, Buchanan2020IntersectionalCH, Kalichman2021FindingTR, Bauer2021LatentVA, Hampton2021BlackFM, Bowleg2021EvolvingIW, schlesinger2017interhci, Rankin2020IntersectionalityIH}. Furthermore, AI ethics researchers have addressed the narrow perspective of intersectionality as intersectional subgroup fairness (e.g., \citet{P3}); our review points to this too, although our scope is wider and considers numerous gaps in AI's operationalization of intersectionality.

A few papers have reimagined intersectionality in AI \cite{sweeney2014critical, ciston2019imagining, birhane2021injustice}, pushing for intersectional practices to be woven into the full AI pipeline, and arguing for a joint interrogation of culture, technology, and solutionist framings of fairness (e.g., critical technocultural discourse analysis \cite{Brock2018CriticalTD}). \citet{2020IntroductionD} illuminates the lack of critical praxis in AI, drawing upon \citeauthor{Collins2017BlackFT}'s matrix of domination to encourage researchers to reflect on how AI relates to ``domination and resistance at each of these three levels (personal, community, and institutional)'' \cite{Collins2017BlackFT}. \citet{P16}, inspired by \citet{Crenshaw_1991}, argue for AI to be reparative and aware of social and historical context. \citet{klumbyte2022critical} facilitate community-based critical analysis of the ``tensions and possibilities'' of integrating intersectional knowledge into machine learning systems. With a shared goal of intersectional AI, we complementarily gauge the epistemic alignment of AI papers related to intersectionality with \citeauthor{Collins2017BlackFT}' intersectionality tenets \cite{collins2020intersectionality}. We go beyond the scope of papers like \citet{Birhane2022FM}, which is not explicitly about intersectionality and focuses on evaluating discussion of social context and power. 

\vspace{-0.2cm}
\section{Intersectionality Overview}
\label{sec:intersect-overview}
Crenshaw coined the term ``intersectionality'' in her 1981 paper \cite{crenshaw1989demarginalizing}, and expanded on it in \cite{Crenshaw_1991}. In the context of violence against Black women, these works study the interactions of race and gender, as well as racism and patriarchy as systems of subordination. Her work is grounded in ``a bottom-up commitment'' to address the needs of those who are ``victimized by the interplay of numerous factors,'' with the explicit goal of obtaining political and social justice. Thus,
praxis has been an important facet of intersectionality from its inception; what constitutes praxis is broad and contextual, including ``movements for economic justice, legal and policy advocacy, state-targeted movements for prison abolition'' \cite{Cho_Crenshaw_McCall_2013}.

While various definitions of intersectionality have emerged, they all center a need to examine power relations across structures, disciplines, domains, and location \cite{Hancock_2007, The_Combahee_River_Collective_1978, Alexander-Floyd_2012}. We draw upon broad intersectionality scholarship in our paper to enrich our own observations. To ground our review methodology and analysis in the following sections, we base our evaluations on \citet{collins2020intersectionality}. This work details six core tenets of intersectionality (drawing from an in-depth genealogy of intersectionality) that lend themselves to an \textit{analytical language} and \textit{cognitive organization} around how forms of oppression are co-created, operated, amplified, and interact with social and structural disparities. These tenets are: social justice, social inequality, relationality, social power, social context, and complexity. We describe each tenet below, its connections to AI fairness, and how we interpret the tenet for advancing social justice in AI fairness.
These descriptions further inform the construction of 3-4 guiding questions per tenet to assess how well the works in our critical review engage with the \textbf{tenets} (Table \ref{tbl:guiding_questions}).

\textbf{Social Justice. }
\label{sec:tenet-justice}
Intersectionality emerges as a synergy between inquiry and praxis, where praxis is action to advance social justice that is informed by inequities identified via critical inquiry (e.g., via the tenets). \citet{collins2020intersectionality} caution that inquiry alone does not further social justice; intersectionality ``demands more than simply being critical and entails turning critical analyses into critical praxis'' \cite{collins2020intersectionality}. In AI fairness, social justice praxis spans numerous practical approaches to fairness, e.g., debiasing techniques, fairness metrics for multiplicative groups; however, its effectiveness depends on authors' social context. Intersectionality widens these practical approaches; this does not remove researchers from the AI fairness domain, but rather deepens our ability to engage with the domain. Overall, intersectionality enables the creation of new forms of knowledge which are informed by a critical examination of how AI systems reproduce inequalities. Therefore, our social justice guiding questions assess how works commit to advancing justice and center the perspectives of subordinated communities.

\textbf{Social Inequality. }
\label{sec:tenet-inequality}
Intersectionality rejects the inevitability of inequality as ``hardwired into the social world, into individual nature'' \cite{collins2019intersectionality}; rather, the framework emphasizes the study of  how social inequalities are fundamentally formed and reinforced through saturated centers of power. Dismantling inequalities requires locating these centers. In AI fairness, inequality is often measured via quantities like demographic parity and disparate impact \cite{Jacobs2019MeasurementAF}. Hence, these metrics ground the practice of harm reduction; however, static measures pointing towards equality rather than equity do not resolve complex and wide-reaching inequality. Instead, intersectionality asks us to center the social and historical context of those at the margins to inform praxis. As such, our inequality guiding questions assess the depth with which researchers situate their work in social inequality.

\textbf{Relationality. }
\label{sec:tenet-relations}
Relationality enables us to examine power and inequality by centering relational thinking. This functions to unveil how concentrations of power take shape, are situated in a broader social context, and perpetuate inequalities. Relationality comprises: addition (what happens when we \textit{don't} consider the intersections of social categories), articulation (how relations impact the growth or dissolution of such intersections), and co-formation (e.g., of social categories as phenomena) \cite{collins2020intersectionality}. In the context of AI fairness, relationality involves examining the relations between decisions we make as researchers, the technical artifacts we produce, and whom they impact (e.g., how the Eurocentrism of auditing frameworks makes them fail to capture inequalities in globally-deployed AI). Hence, our relationality guiding questions assess works' intention and inquiry across technological structures and social context.

\textbf{Social Power. }
\label{sec:tenet-power}
Intersectionality uses relationality to tie ``intersecting power relations'' to how power ``produce[s] social divisions of race, class, gender, etc.'' \cite{collins2020intersectionality}. Intersectionality is predicated on understanding that systems of power ``co-produce each other in ways that reproduce unequal material outcomes and the distinctive social experiences [within] hierachies'' \cite{collins2019intersectionality}. In AI fairness, power is concentrated in \textit{human} choices: system design, data collection, deployment, operationalizations of fairness. These choices impact resource allocation for communities at the intersections of the ``structural, disciplinary, cultural, and interpersonal'' domains \cite{Barocas2016BigDD, Collins2017BlackFT, 2020IntroductionD}; thus, power should be discussed at all stages of the AI pipeline. Our power guiding questions therefore assess the extent to which researchers reflexively comment on or situate their work in the power relations in which they participate.

\textbf{Social Context. }
\label{sec:tenet-context}
Intersectionality centers ``context-specific [...] historical particularities and the increasing significance of a global context'' \cite{collins2020intersectionality}. When engaging with intersectionality in different (especially global) contexts, inquiry and praxis take different forms; consequently, one must practice epistemic, personal, and critical reflexivity to be cognizant of context, in order to effectively and holistically advance justice. In AI fairness, social context informs AI context through researcher training and background, model training and deployment, language choices, etc. Hence, self-reflexively acknowledging that one operates in the Global North informs \textit{who} is centered in fairness tasks. Conversely, fairness works that flatten social context (e.g.,  by optimizing for ``Indigenous people'' broadly) informs who drives knowledge production. As a result, our social context guiding questions assess the extent to which context is deliberately referenced and informs research processes.

\textbf{Complexity. }
\label{sec:tenet-complexity}
Complexity is key to a ``creative tension'' between critical inquiry and praxis, which results in new forms of social action to combat inequality \cite{collins2020intersectionality}. Complexity necessitates relational thinking and situational awareness. In AI fairness, complexity is often conceptualized as minimizing unfairness across a large number of social groups. However, complexity is more expansive; for example, it entails co-designing with groups who have been harmed by AI systems rather than using preconceptions of excluded groups to remedy exclusion. Our complexity guiding questions probe how works contend with model requirements, community needs, and centers of power that influence AI design. This notion of complexity is distinct from how complexity is used in the complex systems discipline, or runtime complexity in CS.

\section{Critical Review Methodology}
\label{sec:review methods}

\subsection{Paper Inclusion Criteria}
To gauge how AI fairness research conceptualizes and operationalizes intersectionality, we curate 30 papers by: 1) querying ``intersectionality machine learning'' on Google Scholar to obtain 75 relevant papers, and 2) filtering those to papers published in AI venues including symposiums, conferences, journals, and books. We choose to query ``machine learning'' as AI fairness research tends to center machine learning. Our process simulates how researchers might discover AI fairness literature related to intersectionality when grounding their own work. Papers are tagged as including intersectionality if they cite intersectionality scholarship that centers critical inquiry. We restrict our sample to 30 papers to ensure that we can annotate each paper (some papers by multiple authors) for engagement with intersectionality. We document all the papers we review in Tables \ref{tbl:meth_tags} and \ref{tbl:inter_tags}, and provide statistics thereof in Table \ref{tbl:review_stats}.

\subsection{Review Methods}
Our annotation scheme is based on the tenets and corresponding guiding questions discussed in \S\ref{sec:intersect-overview}. All questions reflect three  axes of reflexivity: epistemological, personal, and critical \cite{palaganas2017reflexivity}. For each paper, for each guiding question (e.g., ``Do the authors mention power?''), we annotate whether or not the authors of the paper explicitly or implicitly answer the question. Then, for each tenet, we annotate that the paper has characteristics of the tenet if it explicitly or implicitly answers at least one of the guiding questions corresponding to the tenet. Importantly, our questions are not a checklist to determine whether researchers have ``truly'' engaged with intersectionality; rather, they reveal where efforts in AI fairness are concentrated and help us reimagine our practices towards advancing social justice in AI. We share all our guiding questions in Table \ref{tbl:guiding_questions}. We further break down our methodology for creating questions in Appendix \S\ref{app: questions}.

11 out of the 30 papers were evaluated by 3 annotators, and we present our tenet-level interannotator agreement for these papers in Table~\ref{tbl:inter_annot}. The scores in Table~\ref{tbl:inter_annot} indicate moderate to high interannotator agreement. The remaining 19 papers were each evaluated by at least 1 annotator. We expand on our annotation methodology in Appendix \S\ref{app:annot_meth} and provide our annotations at \url{https://tinyurl.com/intersectionality-annotations}.

Given the nature of intersectionality, engagement therewith cannot be captured solely through quantitative means; therefore, we also qualitatively mine intersectionality-related themes from our sample of papers. With these deductive (i.e., using our guiding questions) and inductive (i.e., qualitative coding) analyses, we supply a bird's eye and granular view of engagement with intersectionality in AI fairness. As praxis, we translate our inductive findings to recommendations for deeper engagement with intersectionality. These recommendations are tailored for AI fairness researchers with any level of training in AI, in academia, industry, or both. We urge readers to take their own identity, capacity, and power into account when considering our recommendations, as these will affect what they can do and potential consequences.

In our analyses, we acknowledge that papers are products of varied epistemological contributions, relations between authors and reviewers, and power dynamics. Thus, our critical review is not so much a criticism of AI fairness researchers as it is a reflection of broader systems, such as the incentives and infrastructural forces that govern publishing in CS and enacting change in corporations, as well as the types of knowledge production that are valued or even simply considered legitimate in the field. Papers do not reflect everything that goes into a research project, and they are also merely static snapshots in time that researchers grow beyond.


\subsection{Investigating Intersectionality Within the AI Fairness Research Paradigm}
\label{sec:intersect_ai_research_paradigm}

Reflexivity enables AI fairness researchers to engage in praxis; as \citet{mohamed2020decolonial} comment, ``deciding what counts as valid knowledge, what is included within a dataset, and what is ignored and unquestioned, is a form of power [...] that cannot be left unacknowledged.'' To interrogate knowledge and inspire reflexivity, we texture our deductive analysis of intersectionality in AI fairness via four methodology lenses: where intersectionality is situated in the AI development process, how papers describe sources of bias, types of CS papers, and (inter)disciplinary relationality (i.e., synergy). These methodologies speak to both the research process and structures which researchers navigate in their work. We document the methodology tags for all the papers we review in Table \ref{tbl:meth_tags}.
\subsubsection{Operationalization of intersectionality}
\label{sec:method_op_intersect}
We observe how papers engage with and operationalize intersectionality in the AI pipeline. Papers are tagged as pre-processing (i.e., pre-training interventions), in-processing (i.e., training-time modeling choices), post-processing (i.e., test-time interventions of model predictions), full pipeline, or processes. ``Full pipeline'' situates intersectionality (for empirical work) across the pipeline, while ``processes'' situates intersectionality in broader AI design and epistemology. Works that deeply engage with intersectionality exercise its tenets at every stage of the pipeline. Researchers can contrast modes of operationalizing intersectionality and in that tension, reimagine how they engage with the framework.

\subsubsection{Source of bias}
\label{sec:method_source_bias}
A paper may characterize bias as systemic, statistical, both systemic and statistical, or entirely fail to describe its source. Understanding sources of bias is pivotal to aligning AI fairness with intersectional praxis. Intersectionality posits that unequal outcomes reflect a systemic reproduction of existing power relations \cite{collins2019intersectionality}. Systemic descriptions of bias concern structures and oppressive forces which subsequently permeate sociotechnical systems. In contrast, statistical descriptions limit sources of bias to the model or data. 

\subsubsection{CS paper type}
\label{sec:method_cs_paper}
We study paper types considered valid in CS (as determined by those in positions of power), exposing tensions between intersectionality and visibility, and allowing us to interrogate assumptions about supposed barriers to knowledge due to disciplinary divides \cite{raji2021exclusionary}. We classify papers as theoretical, engineering, empirical, or a combination of types based on \citet{howtorea21:online}. Papers that do not fit any of these types are tagged as ``other.'' This information enables AI fairness researchers to interrogate possible interplays between intersectionality and their epistemology.

\begin{table}[!t]
\centering
\caption{Critical review statistics ($N=30$)}
\vspace{-0.28cm}
\label{tbl:review_stats}
\begin{tabular}{p{6cm}ll} 
\hline
Characteristic                                                       & N  & \%  \\ 
\hline
Intersectionality literature referenced                              & 22 & 0.73  \\
No. papers for annotator agreement                                   & 11 & 0.37  \\
\textbf{Terminology}                                                 &    &       \\ 
\hline
Uses term ``intersectionality''                               & 26 & 0.87  \\
Uses term ``intersectional''                                   & 27 & 0.9   \\
\textbf{AI Pipeline Stage} &    &       \\ 
\hline
Pre-processing                                                       & 5  & 0.17  \\
In-processing                                                        & 4  & 0.13  \\
Post-processing                                                      & 10 & 0.33  \\
Full pipeline                                                        & 5  & 0.17  \\
Processes                                                            & 10 & 0.33  \\
\textbf{CS Research Paradigm}                          &    &       \\ 
\hline
Theoretical                                                          & 10 & 0.33  \\
Empirical                                                            & 23 & 0.77  \\
Engineering                                                          & 11 & 0.37  \\
Other                                                                & 6  & 0.2   \\
Synergy across disciplines                                           & 16 & 0.53  \\
\hline
\end{tabular}
\vspace{-0.2cm}
\end{table}

\subsubsection{Synergy across disciplines}
\label{sec:method_synergy}

Papers are tagged for synergy if they incorporate literature beyond other AI papers and intersectionality scholarship (tagging process described in \S\ref{app: tagging}). By incorporating knowledge forms beyond CS, we make room for dialogue across ``more than one way of knowing'' \cite{smith2021decolonizing, Gebru_2021}. This is particularly important for sources of marginalized knowledge that may go unheard.
\citet{smith2021decolonizing} asserts that knowledge is always situated; dominant academic AI epistemologies describe systems as ``universal'' or ``neutral,'' when in fact these terms simply indicate that other ways of knowing have been subjugated.
Engaging in participatory AI research is one way of ``recovering [...] stories of the past'' \cite{thambinathan2021decolonizing}. However, researchers can also embrace synergy across disciplines. This allows us to examine how AI epistemology's alignment with other works interacts with intersectionality to create new forms of knowledge production towards advancing AI fairness.

\section{Deductive Analysis}
\label{sec: deductive}

\begin{figure}[t]
\centering
\includegraphics[width=.45\textwidth]{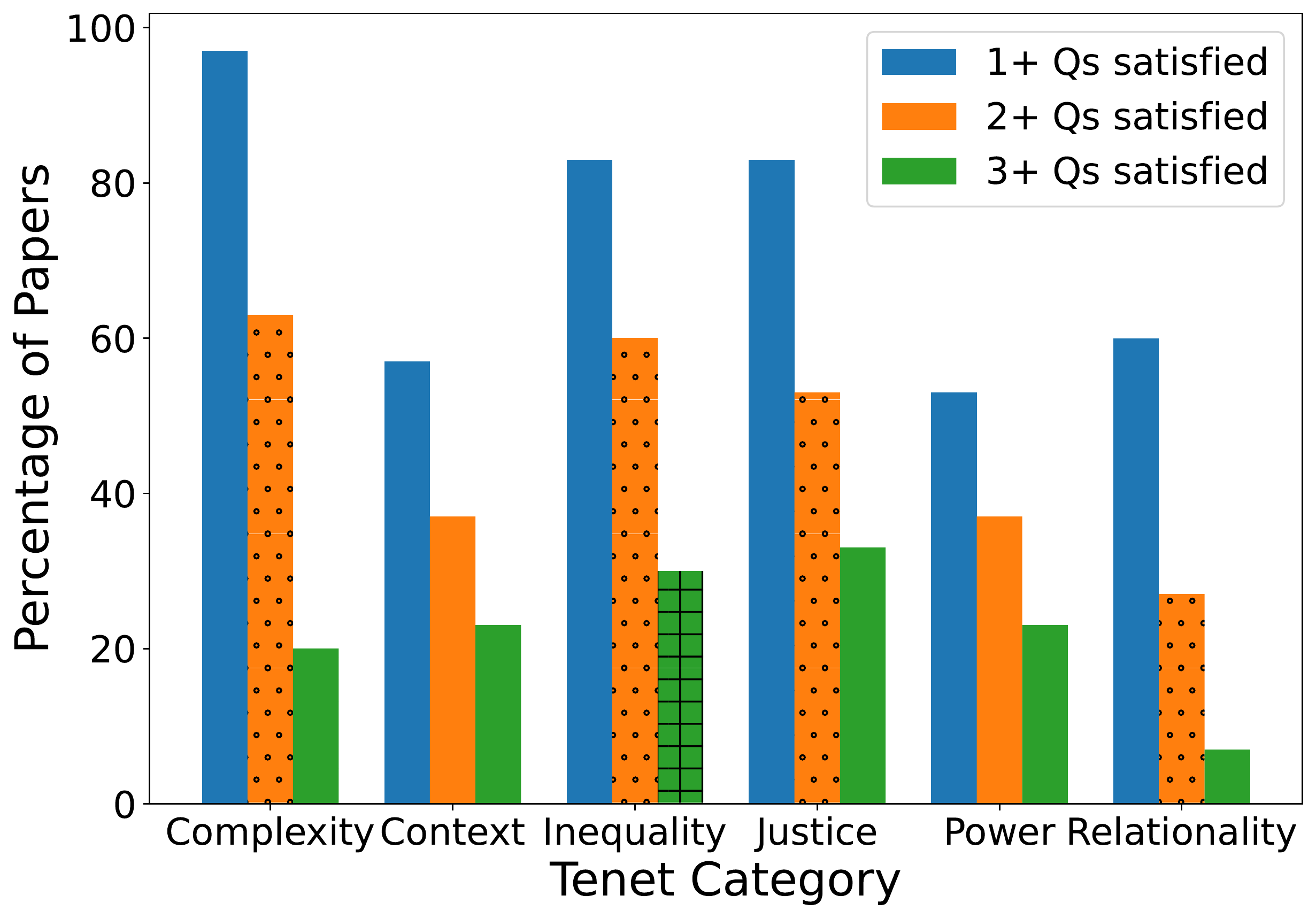}
\vspace{-0.4cm}
\caption{Distribution of intersectionality tenets split by depth of engagement with our guiding questions.}
\label{fig:tenet-bars}
\end{figure}

\begin{figure*}[h]
\begin{subfigure}[t][][c]{0.24\textwidth}
\includegraphics[width=\columnwidth]{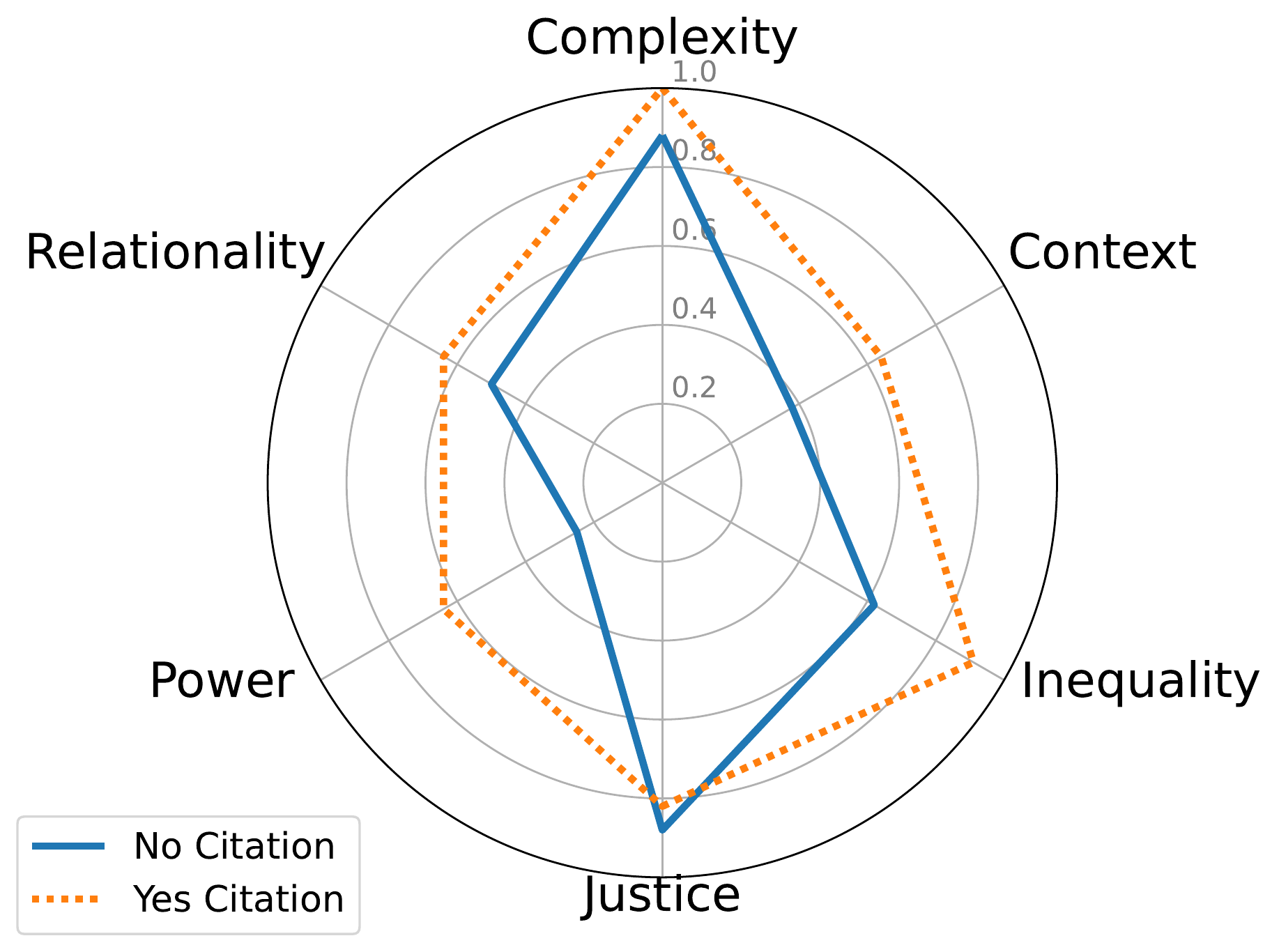}
    \caption{Cites Intersectionality Lit.}
    \label{fig:radar_citation}
  \end{subfigure}
  \begin{subfigure}[t][][c]{0.24\textwidth}
    \includegraphics[width=\columnwidth]{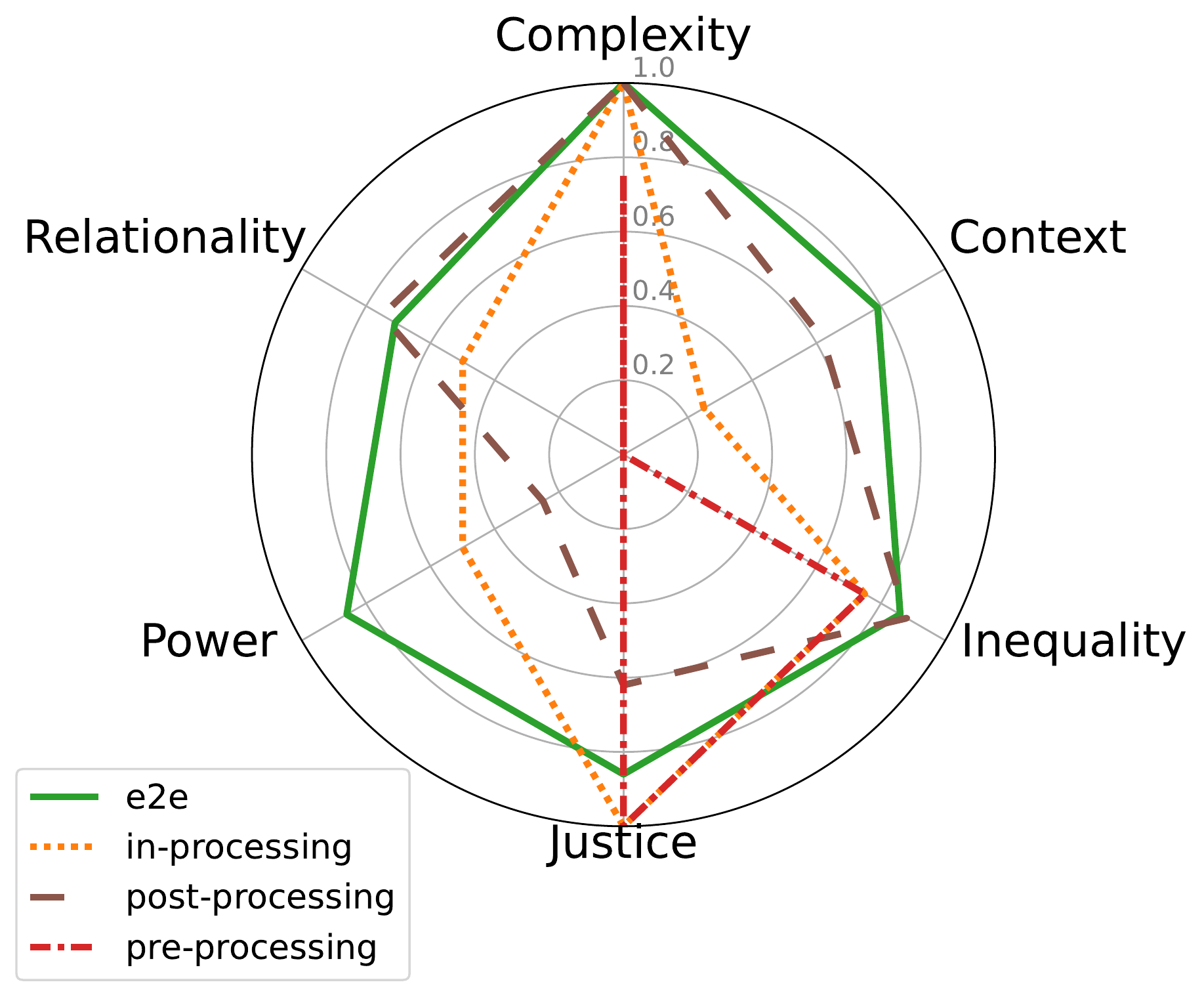}
    \caption{Operationalization}
    \label{fig:radar_intersect}
  \end{subfigure}
    \begin{subfigure}[t][][c]{0.24\textwidth}
    \includegraphics[width=\columnwidth]{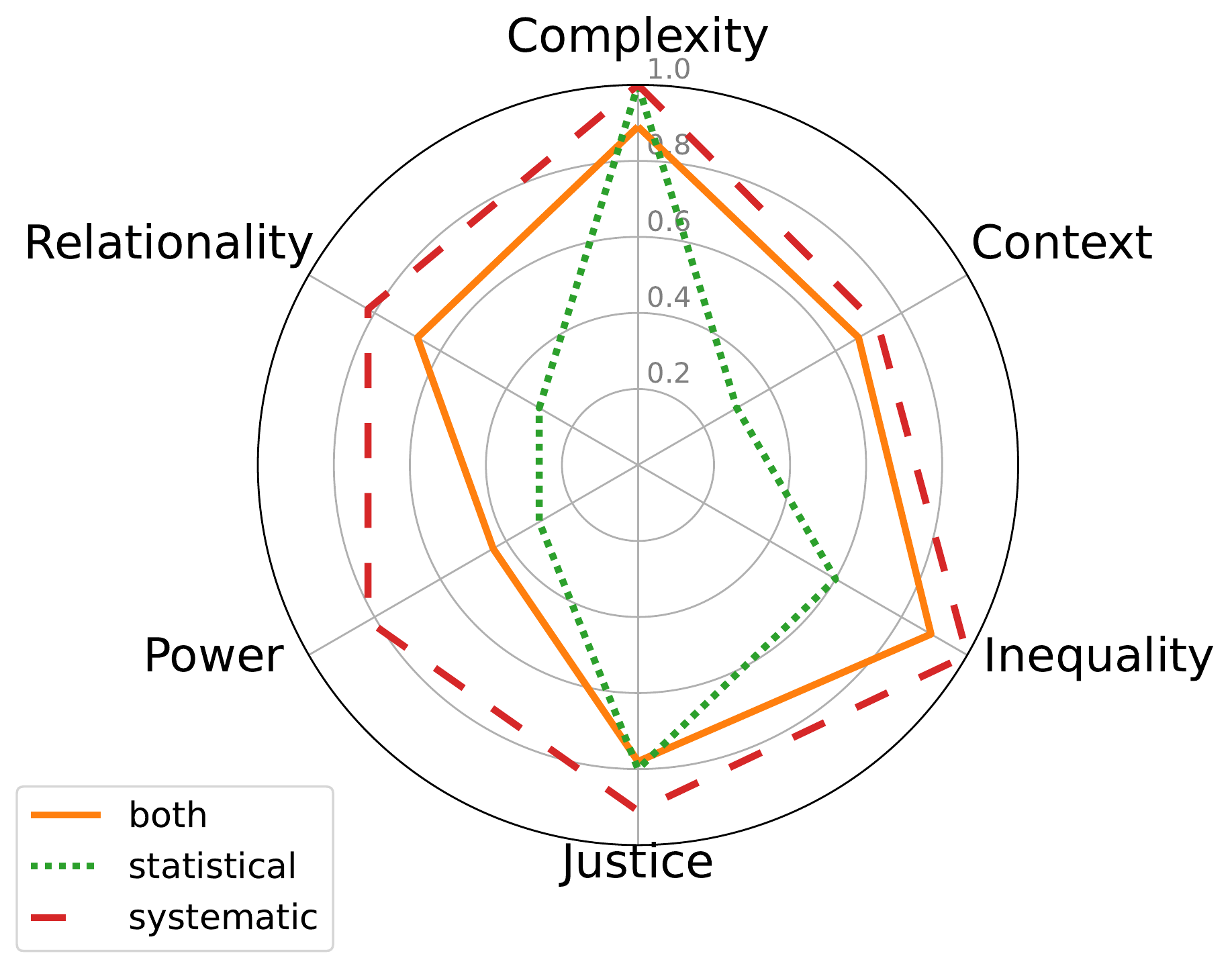}
    \caption{Source of Bias}
    \label{fig:radar_bias}
  \end{subfigure}
    \begin{subfigure}[t][][c]{0.24\textwidth}
    \includegraphics[width=\columnwidth]{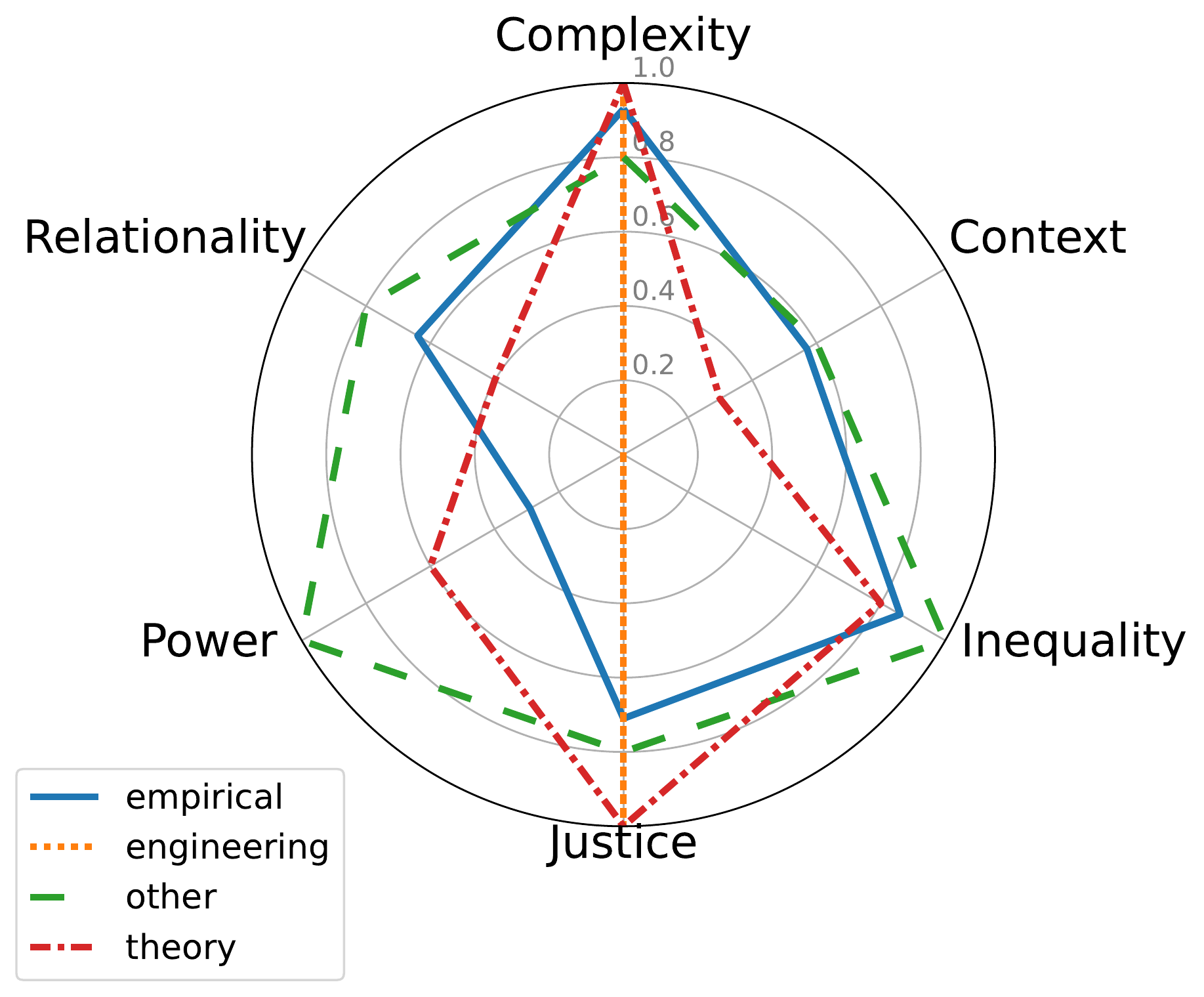}
    \caption{CS Paper Type}
    \label{fig:radar_paper}
  \end{subfigure}
  \vspace{-0.3cm}
  \caption{Relative distributions of papers across intersetionality tenets if a paper engages with at least 1 question per tenet.}
\end{figure*}

\noindent \textbf{Quantitative Summary. }
\label{sec: deductive summary}
We report tenet distributions across all papers in Figure \ref{fig:tenet-bars}. Complexity (97\% of all papers), inequality (83\%), and justice (83\%) appeared most often in works that engaged with at least 1 guiding question. In contrast, the tenets that appeared least often were power (53\%), context (57\%), and relationality (60\%). Taking the number of questions answered as a proxy for depth of engagement with a tenet, we see drops in every tenet. The largest drop (20\%) between answering 1+ questions versus 3+ questions is in complexity, despite high overall engagement. Relationality similarly drops from 60\% to just 7\%. These results are interrelated just as the tenets are; understanding power across structures requires understanding social context and the relations between social groups \cite{Collins2019IntersectionalityAC}. Therefore, it is suspect that a majority of papers purportedly center social justice and inequality with so few discussing power. \\

\noindent \textbf{Cites Intersectionality Literature. }
\label{sec: has citation}
Figure \ref{fig:radar_citation} shows that citation of intersectionality literature (see \S\ref{app: tagging} for more details) affects how papers engage with power, inequality, and context. It does not, however, seem to cause differential engagement with complexity and justice.

64\% of papers that cite intersectionality literature engage with power, compared to 25\% of papers that don't cite it. Engagement with the literature would explain this, as intersectionality is grounded in an analysis of power. However, the overall consideration of power is middling, echoing intersectionality theorists' observation that the ``recasting of intersectionality as a theory primarily fascinated with the infinite combinations and implications of overlapping identities from an analytic initially concerned with structures of power and exclusion is curious given the explicit references to structures that appear in much of the early work'' \citep{Cho_Crenshaw_McCall_2013}.

We see a similarly large gap in engagement with inequality. 91\% of papers that cite intersectionality literature also discuss social inequality as a phenomenon with social and historical roots, or something their work impacts, compared to only 62\% of papers that don't cite it. We see this difference as a reflection of intersectionality's motivation as a framework to examine inequalities.

Papers only seem to show consistent engagement with the tenets of complexity and justice, regardless of whether they cite intersectionality literature (above 80\% of papers in each of these splits). This reflects the CS paradigm of understanding intersectionality as rejecting single axes of identity, and the ethos of AI fairness -- one that seeks justice and a better future. Overall, citing intersectionality literature correlates with deeper tenet engagement.\\

\noindent \textbf{Operationalization of Intersectionality. }
\label{sec: op_intersect}
Figure \ref{fig:radar_intersect} shows differences in how intersectionality is used across the AI pipeline. Papers operationalizing intersectionality end-to-end had the largest coverage across intersectionality tenets, with each tenet appearing in 71-100\% of these papers. Meanwhile, the lowest engagement across tenets came from papers focused on pre-processing, with \textit{none} of them engaging with context, power and relationality.

The locus of operationalization seemed to make the biggest difference in how context and power were engaged with. Engagement with the social context tenet seemed to increase as papers went further down the AI pipeline; no pre-processing focused papers engaged with it compared to 25\% of in-processing focused papers, 50\% of post-processing focused papers, and 71\% of end-to-end papers. This pattern mostly held for power as well, except that in-processing papers (50\%) engaged with this tenet more than post-processing papers (25\%). Overall, papers engage with more tenets when they operationalize intersectionality end-to-end and in processes. \\

\noindent \textbf{Source of Bias. }
\label{sec: source of bias}
Differences in tenet engagement across the source of bias are shown in Figure \ref{fig:radar_bias}. Papers treating the source of bias as statistical had the lowest engagement across tenets, with only 30\% of these papers engaging with context, relationality, and power. On the other hand, these papers have 100\% coverage of the complexity tenet. This could be attributed to a common narrow reading of intersectionality as just multiplying identity categories rather than as a structural analysis or a political critique \citep{Guidroz_Berger}.

When considering bias to be systemic rather than statistical, tenet coverage increases noticeably; engagement with relationality jumps from 30\% to 67\% of papers in the category, context goes from 30\% to 73\%, and inequality goes from 60\% to 100\%. This aligns with existing literature in which discussing the social reality of a phenomenon allows one to more deeply assess the factors that contribute to it in the first place \cite{benjamin2019race}.

Papers that conceive of bias as \textit{both} statistical and systemic have the best tenet coverage overall, with roughly 80-90\% of papers discussing each of complexity, inequality, and justice. This dual conception of bias incorporates both the social and technical aspects of AI systems and how they  may inform or magnify each other. \\

\begin{figure*}[h]
\centering
\begin{subfigure}[t][][c]{0.25\textwidth}
\includegraphics[width=\columnwidth]{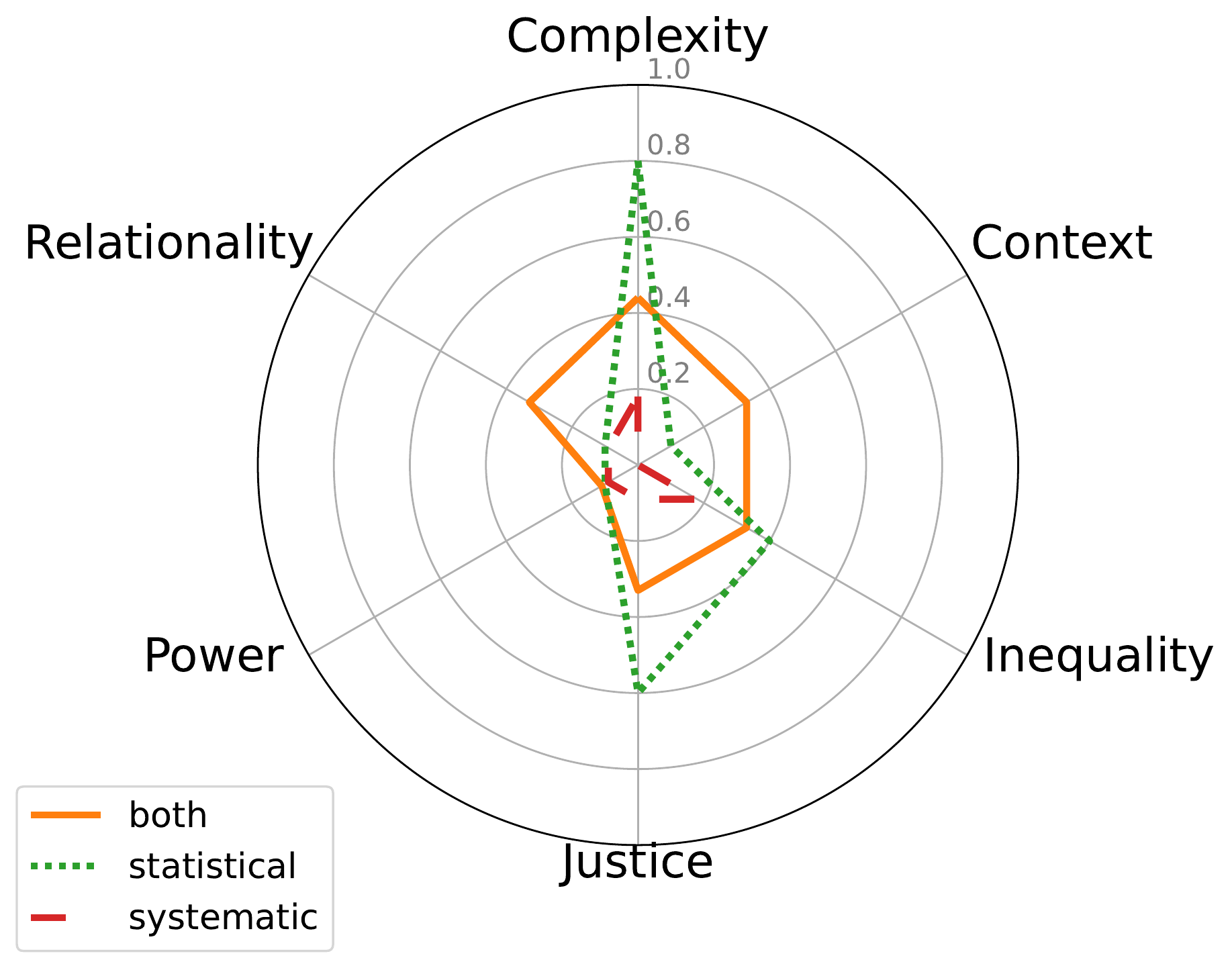}
\end{subfigure}
\begin{subfigure}[t][][c]{0.236\textwidth}
\includegraphics[width=\columnwidth]{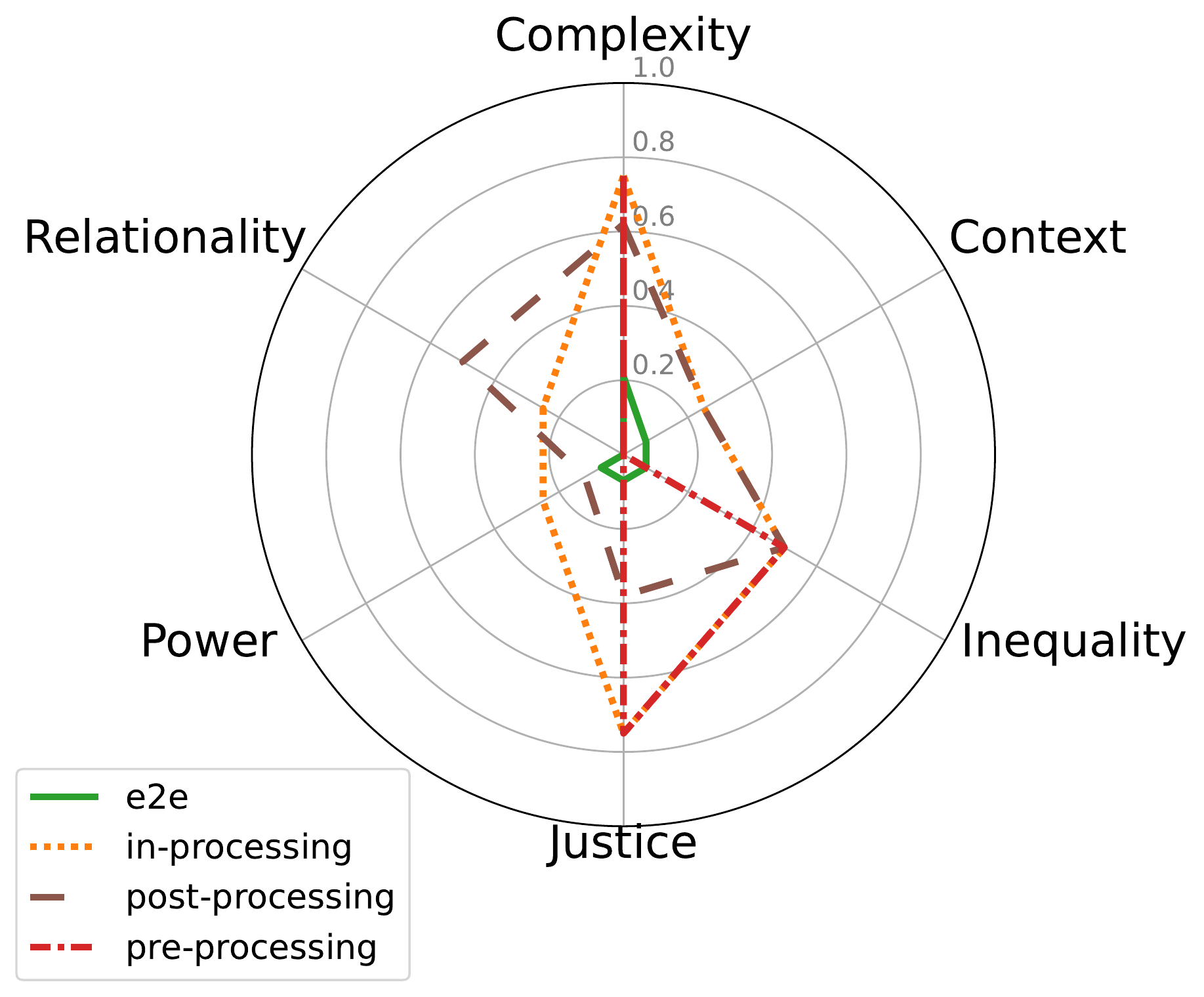}
\end{subfigure}
\begin{subfigure}[t][][c]{0.25\textwidth}
\includegraphics[width=\columnwidth]{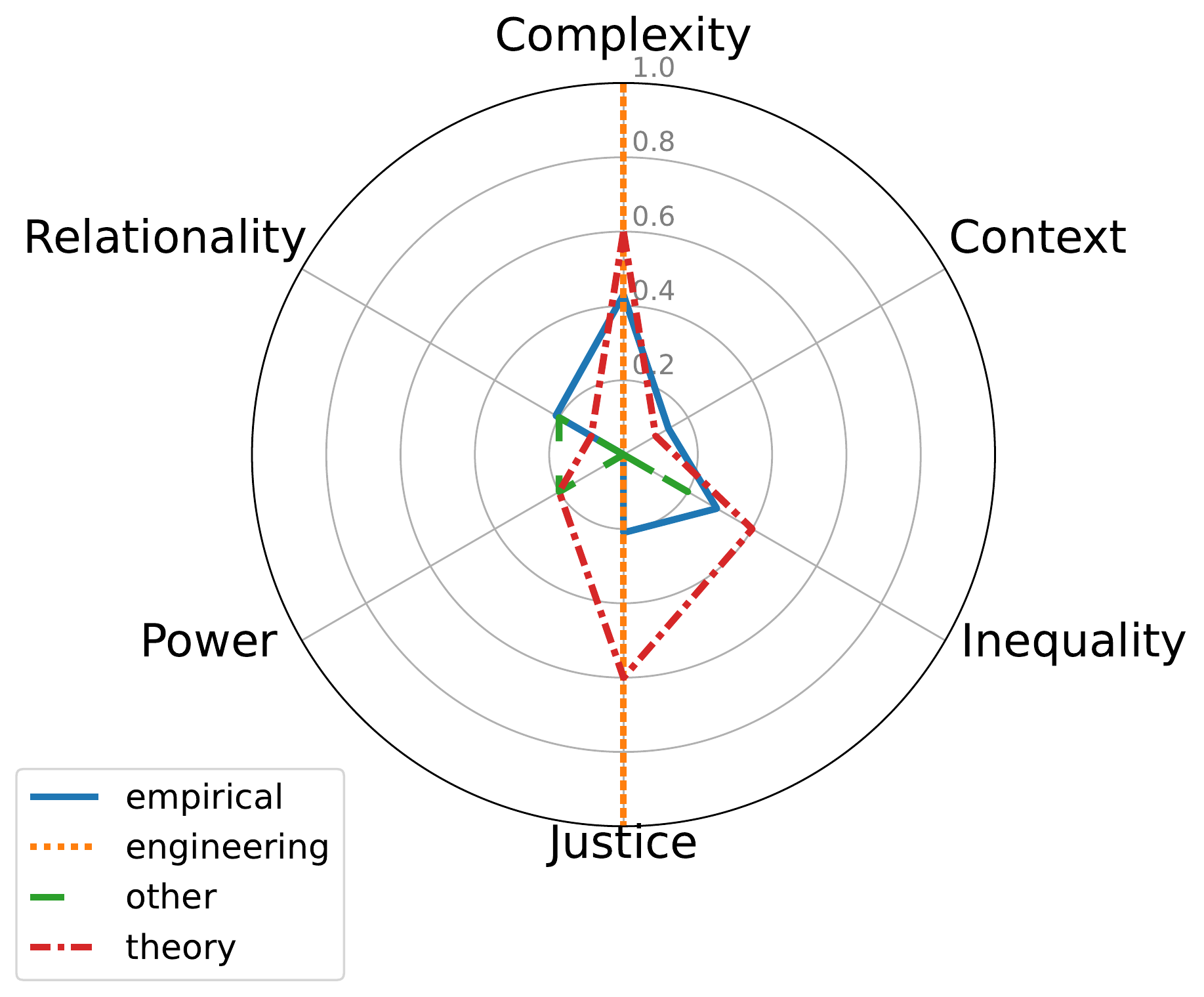}
\end{subfigure}

\begin{subfigure}[b][][c]{0.25\textwidth}
\includegraphics[width=\columnwidth]{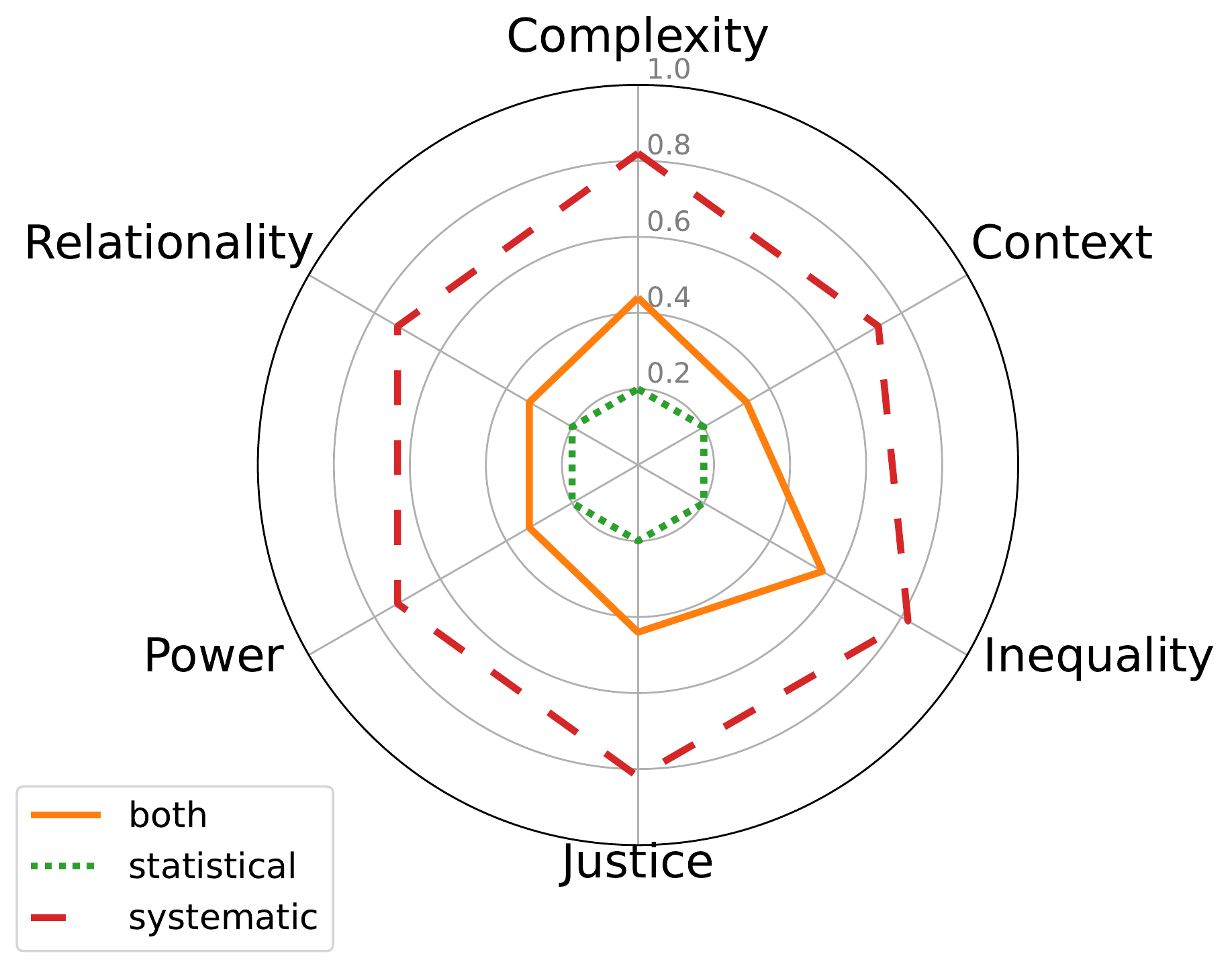}
\caption{}
\label{fig:bias_syn_radar}
\end{subfigure}
\begin{subfigure}[b][][c]{0.236\textwidth}
\includegraphics[width=\columnwidth]{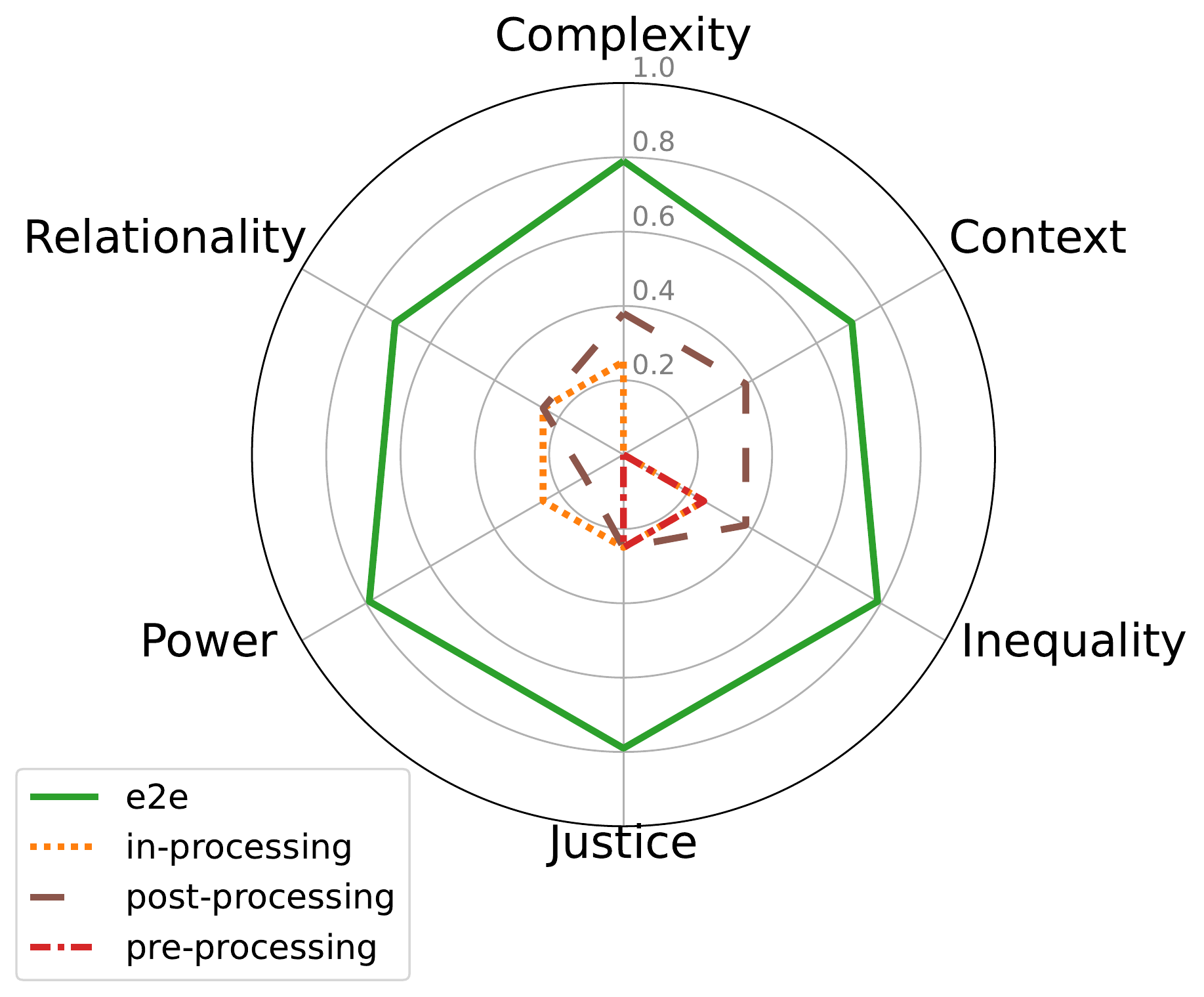}
\caption{}
\label{fig:intersect_syn_radar}
\end{subfigure}
\begin{subfigure}[b][][c]{0.25\textwidth}
\includegraphics[width=\columnwidth]{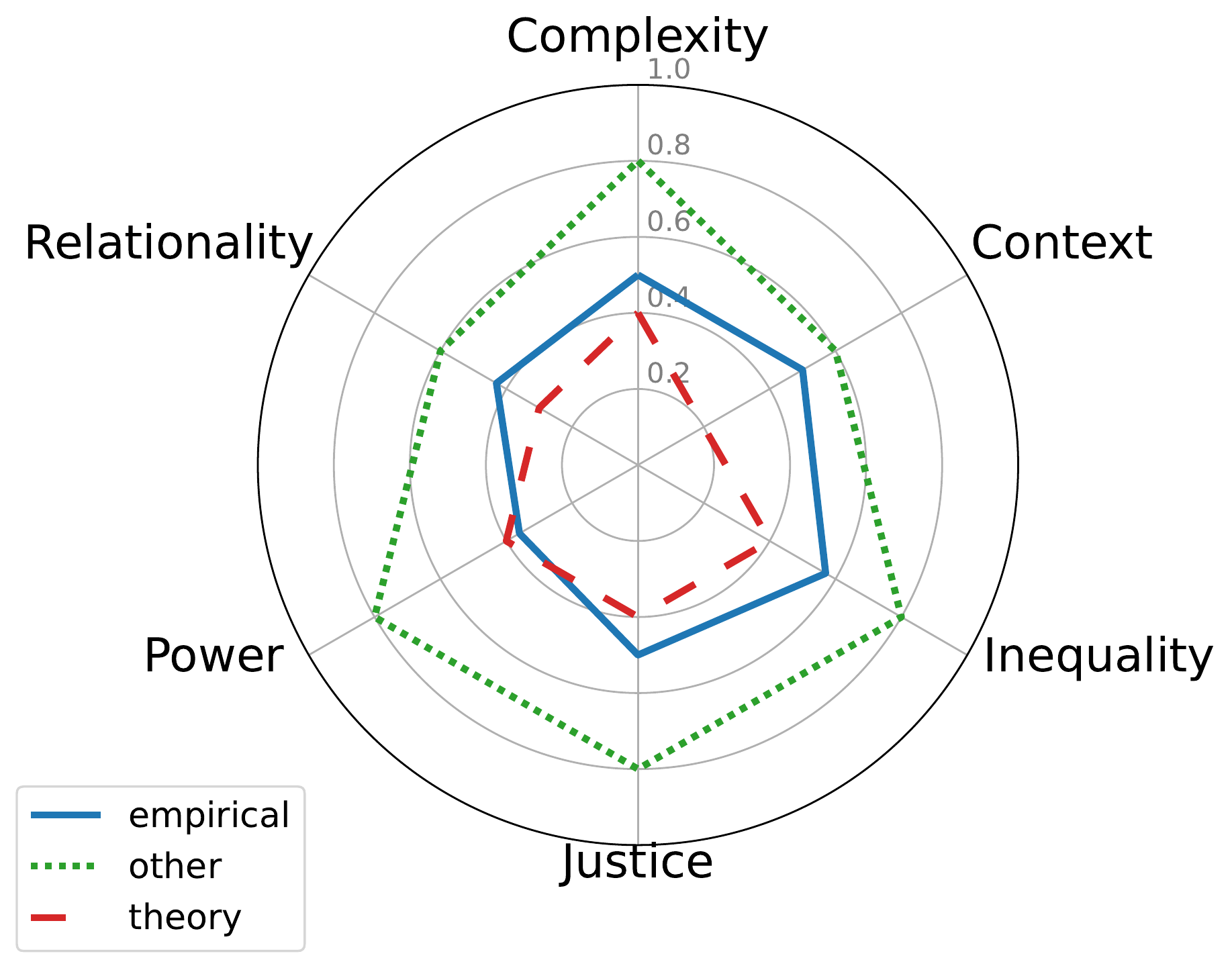}
\caption{}
\label{fig:paper_syn_radar}
\end{subfigure}
\vspace{-0.15cm}
\caption{Papers with at least 1 tenet characteristic, split by presence of a synergistic literary component (top=no, bottom=yes).}
\label{fig:synth-all}
\vspace*{-12pt}\end{figure*}

\noindent \textbf{CS Paper Type.}
\label{sec: cs type}
Figure \ref{fig:radar_paper} shows that papers across all CS paper types consistently engage with complexity and justice, with at least 70\% of papers of each type covering these tenets. This consistency breaks down more dramatically across power, relationality, context, and inequality. \textit{No} engineering papers engaged with these four tenets. At the other end of the spectrum, papers classified as other engaged with the largest array of tenets. 100\% of these papers engaged with power, as opposed to 60\% of theory papers and a quarter of empirical papers. Theoretical papers seemed to engage relatively less with context (32\%) and relationality (41\%). Overall, despite disciplinary divides, papers in CS are able to engage with intersectionality tenets.
Supplementing these findings, our inductive analysis in section \S\ref{sec:subgroup_fairness} indicates that many works use a heuristic definition of intersectionality that is easily operationalized across theoretical, engineering, and empirical papers, resulting in a narrow use of the framework. Engaging with literature outside the empirical and engineering papers that are de rigueur in CS can expand tenet coverage.\\

\noindent \textbf{Synergy Across Disciplines.}
\label{sec: synergy}
Figure \ref{fig:synth-all} shows each tenet category split by whether or not they had a synergistic component. As it pertains to the source of bias, synergistic papers incorporate a wider range of tenets at higher rates than non-synergistic papers (Figure \ref{fig:bias_syn_radar}). Even among papers that treat bias as systemic and thus engage with a social component of bias, tenet coverage benefits hugely from synergy, with 63-73\% more papers discussing complexity and justice. Figure \ref{fig:intersect_syn_radar} shows that when papers that discuss intersectionality across the entire pipeline have a synergistic component, they have better tenet coverage. Synergy appears not to have a big effect on papers that focus on in-processing, pre-processing or post-processing, sometimes even appearing to decrease tenet coverage. With CS paper type (Figure \ref{fig:paper_syn_radar}), papers that incorporated intersectionality in an empirical and theoretical paradigm had better tenet coverage when they had a synergistic component. We note that no engineering papers in our data have a synergistic component -- an interesting finding in its own right.
As before, this suggests that the biggest benefit to tenet coverage can come from first operationalizing intersectionality throughout the pipeline and attending to processes and norms, which arguably \textit{necessitates} interdisciplinary synergy. Overall, disciplinary synergy correlates with higher intersectionality tenet coverage. 

\vspace{-0.15cm}
\section{Inductive Analysis}
\label{sec: inductive}

\textbf{Intersectionality as intersectional subgroup fairness. }
\label{sec:subgroup_fairness}
Among papers which cite intersectionality literature, many conflate  intersectionality with intersectional subgroup fairness. For example, \citet{P14} posit:

\begin{quote}
\textit{``... a model that satisfies conditional parity with respect to race and gender independently may fail to satisfy conditional parity with respect to the conjunction of race and gender. In the social science literature concerns about, potentially discriminated against, sub-demographics are referred to as intersectionality.''}
\end{quote}
\vspace{-0.1cm}
Similarly, \citet{P28} state, ``For intersectional fairness, we created the variable EthnicMarital, engineered by concatenating Ethnic and Marital status.''
Indeed, we observe that many papers conceptualize intersectionality as identity-centric, and its ties to power and inequality are not explicitly named \cite{P25, P26, P28}. Our finding substantiates the concerns of intersectionality scholars that intersectionality is diluted to a ``two-by-two analysis of gender by race'' rather than ``constituting a structural analysis or a political critique'' \cite{Guidroz_Berger}, or contending with ``overlapping systems of subordination'' \cite{Collins2019IntersectionalityAC}. Notably, papers even discuss power and inequality in depth at first, but nevertheless operationalize intersectionality as subgroup fairness without engaging these points again \cite{P5}.

Such additive frameworks are helpful insofar as they enable structural inquiry. However, per our annotations, despite overwhelming discourse on cross-sectional social categories, papers' discussions of subgroups often lack social or historical context \cite{P14, P26, P20}. Few works comment on the structural factors that cause certain groups to be underrepresented in datasets, critically engage with the colonial origins of protected attributes \cite{Fanon1952BlackSW}, or connect groups to social structures and inequality (c.f., the explicit recognition that Black communities are targeted at a higher rate by law enforcement using facial recognition in \cite{P13}).

The obfuscation of intersectionality as subgroup fairness reflects cultural denial, ``the process that allows us to know about cruelty, discrimination, and repression, but never openly acknowledge it'' \cite{Eubanks2018AutomatingIH}. We do not claim that the intentions of AI fairness researchers are malicious; rather, groups ``los[ing] meaning as a descriptive, nonanalytical category'' prevents researchers from engaging in critical inquiry \cite{collins2019intersectionality}. This disarms praxis: AI fairness can no longer contend with advancing justice for those at the margins if their experiences with AI-driven social inequalities are not centered. Therefore, we echo \citeauthor{collins2019intersectionality}' call for ``intellectual vigilance'' in analyzing and articulating intersecting power relations. Using an intersectional lens is crucial to refocusing on marginalized communities, and can inform social justice efforts across various fields by addressing the root causes of harm, regardless of one's training.

\textbf{Recommendation.} Researchers exercise intellectual vigilance when using additive frameworks by creating statistical methodologies that preserve unique social and historical characteristics of intersecting groups. \cite{P1} exemplifies this. Leadership incentivizes this inquiry. Researchers and leadership prioritize widening their conceptualization of intersectionality beyond the ``subgroup fairness'' interpretation, which is limited in its social justice praxis. \\

\vspace{-0.07cm}
\noindent \textbf{Anti-discrimination legislation informs design. }
\label{sec:law}
Several papers draw from regulation (e.g., anti-discriminatory legislation) to define their fairness objective. For example, \citet{P26} state:
\begin{quote}
\textit{``In many---if not most---real-world applications, there are multiple protected attributes (typically 10-20) along which discrimination is prohibited [1, 2].''}
\end{quote}
\vspace{-0.05cm}

\citet{P5} similarly motivate their fairness criteria from a legal perspective: ``consider the 80\% rule, established in the Code of Federal Regulation.''  Additionally, \citet{P2} seek to ``[determine] whether disparities in system behavior meet legal thresholds for discrimination.'' Furthermore, \citet{P15} remark, ``there does not exist a single universally agreed upon definition of fairness,'' citing how different ``anti-discrimination legislation exists in various jurisdictions around the world.''

Motivating AI fairness from a strictly regulatory lens (e.g., the 80\% rule, protected groups) does not fully embrace social and historical context. Several critical scholars argue that discrimination is often legitimized through anti-discrimination law \citep{freeman1977legitimizing, spade2015normal, crenshaw1989demarginalizing, delgado2023critical}. According to \citet{freeman1977legitimizing}, these laws see racial discrimination ``not as conditions, but as actions inflicted on the victim by the perpetrator.'' He adds that such laws reflect the idea that ``only `intentional' discrimination violates anti-discrimination principles,'' creating ``a class of the `innocent' who need not feel any personal responsibility for the conditions associated with discrimination'' \cite{freeman1977legitimizing}. Similarly, in AI fairness, researchers prioritize intersectional subgroup fairness over the structures that give rise to unfairness to begin with. Interestingly, AI fairness researchers who adopt a regulatory lens abundantly cite \citet{Crenshaw_1991}, although this work illuminates how anti-discrimination laws render Black women invisible:

\begin{quote}
\textit{``In a monumental paper published in 1989, Kimberlé Crenshaw [11] introduced Intersectionality by referencing a court case where black women were unfairly discriminated as a result of an activity to mitigate the race and gender discrimination independently'' \cite{P9}.}
\end{quote}

AI fairness researchers must heed the warnings of critical legal studies: indifference to the social and historical context of groups and their intersections risks reproducing histories of discrimination. For example, an important step towards dismantling injustices is challenging social categories rooted in colonialism, as ``this structure imports a descriptive and normative view of society that reinforces the status quo'' \cite{crenshaw1989demarginalizing}; hence, failing to investigate how the mechanisms responsible for the unjust social realities of oppressed groups are upheld by one's technology is fundamentally incompatible with reparation and advancing justice \cite{P16, Cooper2021EmergentUI}. Therefore, while social beneficence may motivate an AI fairness approach, its technical operationalization must consider the sociotechnical environment it operates within. In other words, if the goal of researchers is to leverage intersectionality towards the creation of just AI systems, these systems must be infused with social and historical literacy throughout their lifecycle to prevent indifferent engagement with the people they affect.

\textbf{Recommendation.} Researchers, including those in leadership, critically engage with and remain vigilant of how operationalizations of anti-discrimination laws in their AI systems do not automatically mean that their systems are fair to marginalized communities. They may do this by engaging with critical legal studies texts \cite{freeman1977legitimizing, spade2015normal, crenshaw1989demarginalizing, delgado2023critical} and marginalized communities to learn how they are unfairly impacted even by systems that pass legal audits. \cite{P19} does a good job of examining the tensions between prioritizing different forms of fairness. \\

\enlargethispage{10pt}

\noindent \textbf{Angles of power examined: technodeterminism rules. }
\label{sec:power}
\citeauthor{Collins2017BlackFT} describes intersectionality as examining the mutual influences which ``intersect and interlock'' across ``structural, disciplinary, cultural, and interpersonal'' \cite{Collins2017BlackFT} domains of power. However, among papers that cite intersectionality literature, power is the least engaged tenet, with ``power'' mentioned in only 53\% of papers. Moreover, merely mentioning ``power'' does not entail engaging with it in depth, e.g., \citet{P5} write in their abstract that ``intersectionality [...] analyzes how interlocking systems of power and oppression affect individuals along overlapping dimensions,'' but do not discuss power elsewhere in their paper. Similarly, \citet{P12} only mention power in their related works section. 

Furthermore, across papers that do engage with power and power relations, engagement style varies. For example, we see power described as a distributable commodity; \citet{P7} assert, ``our work [...] stems from the acknowledgment that power is not equally distributed in the world.'' In contrast, \citet{P10} note, without explicitly using the term ``power,'' that ``models can exacerbate existing biases in data and perpetuate stereotypical associations to the harm of marginalized communities.''

One can argue that AI fairness researchers study mechanisms of inequality, namely the way inequalities emerge as ``AI harms,'' so that we may reduce them. As such, the allocational and representational harms of our systems are the result of power enacted by our systems unto those at the margins. We do not reject these approaches to making sense of power discrepancies observed in AI-driven systems. However, many AI fairness researchers constrain their discussion of power to the AI system alone, removing themselves from the equation. The notion that a system itself exerts power is technodeterministic, i.e., it reifies the idea that systems, and not their creators, are responsible for reproducing inequalities. Only a few papers that we review escape technodeterminism, e.g., \citet{P19} state, ``The second alternative perspective focuses on the distribution of power and asks: who gets to pick the objective function of an algorithm? The choice of objective functions is intimately connected with the political economy question of who has ownership and control rights over data and algorithms.'' Engaging with intersectionality forces researchers to shed their technodeterminism and contend with the value-laden choices made by the humans that contribute to the lifecycle of AI systems. This is central to praxis that may effectively advance justice in AI fairness.

\textbf{Recommendation.} Researchers flex intellectual vigilance by being explicit about how their methodologies may contribute to perpetuating social inequalities. They state their full-pipeline design choices at the beginning of projects and iterate as designs are updated. Leadership gives researchers opportunities to engage in critical reflexivity. These issues are further discussed in \citep{benjamin2019race, DIgnazio2021DataFT, Noble_2018}. \\

\noindent \textbf{Questionable citational praxis of intersectionality. }
\label{sec:citational_praxis}
Several papers reference literature incorrectly to justify their operationalization of intersectionality. For example, \citet{P15} assume that \citet{P13} concerns intersectionality though it is actually a study of intersecting subgroups. We see this phenomenon again in \citet{P21}, which cites only \citet{P13} when describing intersectionality. In contrast, some papers, like \citet{P30}, discuss intersectionality, but only cite a paper on affirmative action \cite{kannan2019downstream}. Other papers, like \citet{P5} and \citet{P28}, mention intersectionality, yet do not reference any relevant literature at all; this is reflected in our deductive analysis, with 19\% of papers that use the term ``intersectionality'' not citing any intersectionality literature.

These findings exemplify a weak spot in the citational praxis of AI fairness researchers. \citet{Alexander-Floyd_2012} calls for us to cite intersectionality literature, showing that within social science literature, there has been an erasure of Black women and Black feminist knowledge in papers that discuss intersectionality. She describes the centering of positivist and empiricist methods of knowledge production as a force that (re-)subjugates Black feminist knowledge and contributes to maintaining the status quo of whose knowledge counts as ``scientific'' \cite{Alexander-Floyd_2012}. \citet{Bilge_2013} identifies similar power structures in feminist studies and the broader neoliberal academy that contribute to ``neutralizing the critical potential of intersectionality for social justice-oriented change.''

We find this gentrification of intersectionality in our field too; AI research interprets intersectionality as a dimension of ``solvability'' and scale, ``perpetuat[ing] the status quo injustice'' \cite{P3}. Furthermore, potentially due to disciplinary barriers or gaps, papers use vague language when describing intersectionality. For instance, \citet{P9} assert, ``the concept of Intersectionality covers diverse discussions including the issue of the oppression that people feel due to the discrimination [15].'' \citet{P24} mention ``the complex and interconnected nature of social biases.'' \citet{P18} state, ``an individual’s identity and experiences are shaped [...] by a complex combination of many factors.'' Vague language prevents intersectionality from being appropriately situated in sociotechnical systems, and may convey an incomplete understanding of intersectionality, neutralizing both researchers' and readers' engagement with power structures and inequality.

\textbf{Recommendation.} Researchers explicitly share how their interpretation of intersectionality literature informs their methodology and assumptions. They read critical social justice literature outside of CS and cite it when incorporating it in AI design. Researchers, including those across leadership, expect and enforce intersectionality citational integrity when peer-reviewing. \\

\noindent \textbf{Intersectional AI fairness lacks relationality. }
\label{sec:relationality_varies}
We find that AI fairness researchers have adopted intersectionality in a way that strips the relationship between structures from the complexity of intersectionality’s arguments. This ``misrepresents [the] initial intent'' of intersectionality \cite{Collins2015IntersectionalitysDD, Waseem2021DisembodiedML}, i.e., to question ``how larger social structures influence supposed group level differences'' \cite{Buchanan2020IntersectionalCH}. For instance, some works that engage with intersectionality literature propose statistical solutions for inequality, e.g., \citet{P2} tackle data sparsity by exploiting the structure of data distributions of data-dense subgroups (e.g., white women, Black men) to inform the data distribution of data-sparse subgroups (e.g., Black women). We do not reject statistical approaches to reducing AI harms; however, formulations that do not situate their statistical methods in a social context by, for instance, stating statistical \textit{and} social assumptions those methods are based on, entirely miss the point of intersectionality as a critical framework.

Being intellectually vigilant about the relationship between statistics and the social sciences is crucial for their intersection. However, we observe different levels of contending with this intersection. Vigilance is missing entirely when the assumptions and reasoning behind the translation from social science knowledge to statistics is not explicit (e.g., \cite{P28, P20, P14}), with \citet{P14} describing: ``In the social science literature concerns about, potentially discriminated against, sub-demographics are referred to as intersectionality [12]. More formally, this work proposes a simple approach to ensure group fairness in expectation across an arbitrary set of subgroups.'' \citet{P22} provides a more intentional socio-technical translation: ``although all value combinations are assessed for intersectional fairness, some subgroups may be semantically meaningless and hence should not be returned as the output,'' though what is ``meaningful'' is not described.  Other works go into more depth with their assumptions, (e.g., \cite{P1}, \cite{P18}), with \citet{P18} stating with respect to subgroup formation that ``collaboration with policy, privacy, and legal experts is necessary in order to ascertain which groups may be responsibly inferred, and how that information should be stored and accessed.'' 

We caution against citing intersectionality literature while ignoring the relationships between the structures that create social categories. This fortifies the fallacy that we have engaged in intersectional praxis if we statistically supplement missing knowledge without examining the embedded assumptions and implications of doing so. It is through this neutralization of critical vigilance and reflexivity that AI fairness researchers are unable to identify where social inequalities may emerge through their own praxis. Invoking an intersectional lens enables this and is, therefore, pivotal to understanding the interlocking systems that produce AI injustices and doing AI justice work.

\textbf{Recommendation.} Researchers remain intellectually vigilant about how scholarship from the social sciences relates to and informs both statistical and wider research methodology. As a result, they preserve the social context of social groups when employing statistical methods, e.g., by transparently stating how they infuse statistical assumptions with context. Across points of power, researchers have ``vigilance check-ins'' to check translative asssumptions during AI development milestones. \cite{P18} engages with transparency at the model level which complements these points.\\


\noindent \textbf{AI social justice praxis varies. }
\label{sec:praxis_varies}
Some papers treat improved fairness as social justice praxis regardless of the task's context. For example, \citet{P2} use recidivism prediction as a fairness benchmark task. As recidivism prediction is a ``byproduct of ongoing regimes of selective policing and punishment'' \cite{benjamin2019race}, the task only serves to uphold the carceral state \cite{Hampton2021BlackFM}. Here, intersectionality posits sites of violence are saturated intersections of power \cite{collins2020intersectionality}.  

Furthermore, many works are not grounded in social context, which ought to inform social justice praxis \cite{P6, P9, P11}. Some papers provide context (e.g., data collection is ``biased toward non-minorities'' \cite{P2}), but nevertheless prioritize generalization \cite{P2}. Some papers even give credence to inferring the social category of individuals; \citet{P14} state, ``gender labels were inferred using the employees’ first names, parsed through the gender-guesser python library.'' Furthermore, we identify works that highlight the oppressive nature of social categories though often defer contestations to future work. For example, \citet{P10} advocate:
\vspace{-0.5cm}
\begin{quote}
\textit{``Future research is recommended to make ground truth comparisons across a broader range of countries against the set of gender-intersections examined in this paper and to comment on a broader spectrum of gender identities.''}
\end{quote}
\vspace{-0.2cm}
Moreover, few papers complement technical contributions with social action, and some even tout their ``purely statistical approach'' \cite{P26}; this neglects the complexity inherent to dismantling social injustices. Mathematical saviorism restricts the operationalization of critical praxis to the pre/in/post-processing stages. This encourages AI researchers to locate sources of unfairness situated only within the technical domain, ignoring the broader sociotechnical milieu linked to the power relations and inequalities upheld by AI \cite{DIgnazio2021DataFT}. Consequently, people already at the margins are erased, even in these contexts that ostensibly address fairness, oppression, and complexity. Thus, AI fairness researchers must engage in praxis that is informed by the experiences of those at the margins.

Some papers justify design choices that do not center care for those at the margins through utilitarian perspectives, e.g.,
\citet{P26} reason, ``an algorithm which discriminates 1 person among a 1000 can be described as fair to an extent.'' On the other hand, works like \citet{P7} and \citet{P18} concretely advocate to dismantle injustice and shift power through participation in model development and transparency in deployment, respectively. 
The contrast in social justice praxis is notable; AI fairness researchers must consider how design choices situate AI systems within their sociotechnical context.

As \citet{crenshaw1989demarginalizing} has said, "addressing the needs and problems of those who are most disadvantaged" means that "others who are singularly disadvantaged would also benefit." Centering these people and the contexts tied to their oppression deepens social justice engagement and creates equity. This way of engaging with intersectionality thus equips AI fairness researchers, regardless of training, to better address inequalities and injustices in AI.

\textbf{Recommendation.} Researchers bridge social justice inquiry and praxis by investing in and valuing the knowledge from communities that their AI systems harm. Researchers and leadership make sure that the AI design process prioritizes harm reduction to promote justice for marginalized communities. \cite{P7} does a good job at centering AI development through community engagement.  \\

\noindent \textbf{AI fairness misses critical reflexivity. }
\label{sec: lack reflex}
Several papers neglect to state their social context and its implications for research methodology. This is reflected in our annotations, with 43\% of papers---even critical ones, e.g., \citet{P3}---not including their social context (often the US) \cite{P4, P20, P25}. Furthermore, when describing social context, some works only include the US as an important context, without commenting on the aspects of complexity and power inherent to doing so. This privileges western contexts as the ``default'' context, resulting in western prototypicality (c.f., white prototypicality \cite{Gordon_2006}). For instance, \citet{P10} argue:

\vspace{-0.2cm}
\begin{quote}
\textit{``using US data may provide an appropriate baseline comparison: 50\% of Reddit traffic comes from the US, and a further 7\% from Canada and the UK each [34]. Given that US sources form a majority in GPT-2's training material [...], we consider the US dataset a satisfactory first benchmark.''}
\end{quote}
\vspace{-0.2cm}

Moreover, when authors do name their social context, they often phrase it as a blanket limitation rather than a contextualization of their research choices; \citet{P12} share that ``the social construction and definitions of sensitive attributes'' are ``outside the scope of the present work but which are important in any real application.'' Stating their context as a limitation---instead of a point which textures their work from the onset ---situates their context as an afterthought, rather than something that undergirds the entire research process. On the other hand, \citet{P7} center reflexivity throughout their work stating: ``Throughout this process, we take an explicitly feminist approach, both in our overaching process—which we strive to make iterative, reflexive, contextual, and participatory—as well as the technology we build''.

All in all, critical reflexivity is crucial to operationalizing intersectionality, both as inquiry and praxis. AI researchers are overwhelmingly located in the Global North \cite{Birhane2022FM}, which makes many power relations and AI injustices invisible to us, especially when we lack the abilities to inquire upon it. Reflexivity requires that we observe the power relations we participate in or benefit from, dismantle these relations, and identify opportunities for social justice within AI fairness. Our advice aligns with conceptualizations of decolonization within the computational sciences; \citet{birhane2020towards} comment that decolonizing ``requires the beneficiaries of the current systems to acknowledge their privilege and actively challenge the system that benefits them.'' 

Works that decouple social context and relationality from intersectionality may reflect academic incentives (e.g., conference acceptances, funding \cite{BliliHamelin2022MakingIE}, citations) and infrastructural forces (e.g., conference paper formats, objectivity-washing). These push AI researchers to make ``fairness'' palatable by treating it as a complexity-free scientific quantity that can be optimized \cite{Talat2021DisembodiedML, Birhane2019AlgorithmicIT}. Our paper is bound by similar constraints;  we empirically validate our critical analyses in order to publish and our citation of the papers we review gives them ``academic currency'' even as we critique them.

\textbf{Recommendation.} Researchers across points of power iteratively dialogue on unlearning ``universal'' frameworks of knowledge and remain vigilant of \textit{whose} knowledge is centered when developing AI. Leadership incentivizes and provides resources for their team to engage in critical reflexivity tools throughout development. \cite{P8} provides a good example of iterative reflexivity.

\section{Conclusion}
What we cannot name, we cannot see. What we cannot see, we cannot address. By examining AI fairness papers related to intersectionality, we identify several patterns in how the literature discusses intersectionality and how it impacts our ability to produce equitable tools. While our field has much energy to get this technology right, we caution the community against assuming that surmounting a ``fairness issue'' pre/in/post the AI pipeline means we have fixed the social reality driving the problem. This work does not seek to discard existing AI fairness work; instead, we invite a widening of AI fairness practice by centering marginalized people and valorizing critical knowledge production that makes room for their voices. We provide recommendations grounded in producing critical knowledge on how AI systems reproduce social inequalities. Our recommendations are not mutually exclusive with respect to AI fairness infrastructure. Rather, they empower researchers to flex the intellectual vigilance required to produce intersectional work, regardless of CS paradigm. Expanding both the conceptualization and operationalization of intersectionality will enable AI fairness researchers across points of power to engage in deeper social justice praxis for AI. To do this, we advocate for adopting intersectionality as an analytical sensibility rather than an axis of optimization.

\begin{quote}
\textit{``I lack imagination you say}

\textit{No. I lack language.}
\textit{The language to clarify} \newline
\textit{my resistance to the literate.''}

- \textit{Cherr\'{i}e Moraga (1983)} 
\end{quote}

\begin{acks}
We are immensely grateful for the work of intersectionality scholars, especially Black women scholars. We thank Dr. Lisa Bowleg and the anonymous reviewers for their feedback.
\end{acks}


\bibliographystyle{ACM-Reference-Format}
\bibliography{acmart}


\begin{thebibliography}{96}


\ifx \showCODEN    \undefined \def \showCODEN     #1{\unskip}     \fi
\ifx \showDOI      \undefined \def \showDOI       #1{#1}\fi
\ifx \showISBNx    \undefined \def \showISBNx     #1{\unskip}     \fi
\ifx \showISBNxiii \undefined \def \showISBNxiii  #1{\unskip}     \fi
\ifx \showISSN     \undefined \def \showISSN      #1{\unskip}     \fi
\ifx \showLCCN     \undefined \def \showLCCN      #1{\unskip}     \fi
\ifx \shownote     \undefined \def \shownote      #1{#1}          \fi
\ifx \showarticletitle \undefined \def \showarticletitle #1{#1}   \fi
\ifx \showURL      \undefined \def \showURL       {\relax}        \fi
\providecommand\bibfield[2]{#2}
\providecommand\bibinfo[2]{#2}
\providecommand\natexlab[1]{#1}
\providecommand\showeprint[2][]{arXiv:#2}

\bibitem[Alexander-Floyd(2012)]%
        {Alexander-Floyd_2012}
\bibfield{author}{\bibinfo{person}{Nikol~G. Alexander-Floyd}.}
  \bibinfo{year}{2012}\natexlab{}.
\newblock \showarticletitle{Disappearing Acts: Reclaiming Intersectionality in
  the Social Sciences in a Post-Black Feminist Era}.
\newblock \bibinfo{journal}{\emph{Feminist Formations}} \bibinfo{volume}{24},
  \bibinfo{number}{1} (\bibinfo{year}{2012}), \bibinfo{pages}{1–25}.
\newblock
\showISSN{2151-7371}
\urldef\tempurl%
\url{https://doi.org/10.1353/ff.2012.0003}
\showDOI{\tempurl}


\bibitem[Barocas et~al\mbox{.}(2019)]%
        {barocas-hardt-narayanan}
\bibfield{author}{\bibinfo{person}{Solon Barocas}, \bibinfo{person}{Moritz
  Hardt}, {and} \bibinfo{person}{Arvind Narayanan}.}
  \bibinfo{year}{2019}\natexlab{}.
\newblock \bibinfo{booktitle}{\emph{Fairness and Machine Learning: Limitations
  and Opportunities}}.
\newblock \bibinfo{publisher}{fairmlbook.org}.
\newblock
\newblock
\shownote{\url{http://www.fairmlbook.org}}.


\bibitem[Barocas and Selbst(2016)]%
        {Barocas2016BigDD}
\bibfield{author}{\bibinfo{person}{Solon Barocas} {and}
  \bibinfo{person}{Andrew~D. Selbst}.} \bibinfo{year}{2016}\natexlab{}.
\newblock \showarticletitle{Big Data's Disparate Impact}.
\newblock \bibinfo{journal}{\emph{California Law Review}}
  \bibinfo{volume}{104} (\bibinfo{year}{2016}), \bibinfo{pages}{671}.
\newblock


\bibitem[Bauer et~al\mbox{.}(2021)]%
        {Bauer2021LatentVA}
\bibfield{author}{\bibinfo{person}{Greta~R. Bauer}, \bibinfo{person}{Mayuri
  Mahendran}, \bibinfo{person}{Chantel Walwyn}, {and} \bibinfo{person}{Mostafa
  Shokoohi}.} \bibinfo{year}{2021}\natexlab{}.
\newblock \showarticletitle{Latent variable and clustering methods in
  intersectionality research: systematic review of methods applications}.
\newblock \bibinfo{journal}{\emph{Social Psychiatry and Psychiatric
  Epidemiology}}  \bibinfo{volume}{57} (\bibinfo{year}{2021}),
  \bibinfo{pages}{221--237}.
\newblock


\bibitem[Benjamin(2019)]%
        {benjamin2019race}
\bibfield{author}{\bibinfo{person}{Ruha Benjamin}.}
  \bibinfo{year}{2019}\natexlab{}.
\newblock \showarticletitle{Race after technology: Abolitionist tools for the
  new jim code}.
\newblock \bibinfo{journal}{\emph{Social forces}} (\bibinfo{year}{2019}).
\newblock


\bibitem[Berger and Guidroz(2010)]%
        {berger2010intersectional}
\bibfield{author}{\bibinfo{person}{Michele~Tracy Berger} {and}
  \bibinfo{person}{Kathleen Guidroz}.} \bibinfo{year}{2010}\natexlab{}.
\newblock \bibinfo{booktitle}{\emph{The intersectional approach: Transforming
  the academy through race, class, and gender}}.
\newblock \bibinfo{publisher}{Univ of North Carolina Press}.
\newblock


\bibitem[Bilge(2013)]%
        {Bilge_2013}
\bibfield{author}{\bibinfo{person}{Sirma Bilge}.}
  \bibinfo{year}{2013}\natexlab{}.
\newblock \showarticletitle{Intersectionality Undone: Saving Intersectionality
  from Feminist Intersectionality Studies}.
\newblock \bibinfo{journal}{\emph{Du Bois Review: Social Science Research on
  Race}} \bibinfo{volume}{10}, \bibinfo{number}{2} (\bibinfo{year}{2013}),
  \bibinfo{pages}{405–424}.
\newblock
\showISSN{1742-058X, 1742-0598}
\urldef\tempurl%
\url{https://doi.org/10.1017/S1742058X13000283}
\showDOI{\tempurl}


\bibitem[Binns(2019)]%
        {Binns2019OnTA}
\bibfield{author}{\bibinfo{person}{Reuben Binns}.}
  \bibinfo{year}{2019}\natexlab{}.
\newblock \showarticletitle{On the apparent conflict between individual and
  group fairness}.
\newblock \bibinfo{journal}{\emph{Proceedings of the 2020 Conference on
  Fairness, Accountability, and Transparency}} (\bibinfo{year}{2019}).
\newblock


\bibitem[Birhane(2021)]%
        {birhane2021injustice}
\bibfield{author}{\bibinfo{person}{Abeba Birhane}.}
  \bibinfo{year}{2021}\natexlab{}.
\newblock \showarticletitle{Algorithmic injustice: a relational ethics
  approach}.
\newblock \bibinfo{journal}{\emph{Patterns}} \bibinfo{volume}{2},
  \bibinfo{number}{2} (\bibinfo{year}{2021}), \bibinfo{pages}{100205}.
\newblock
\showISSN{2666-3899}
\urldef\tempurl%
\url{https://doi.org/10.1016/j.patter.2021.100205}
\showDOI{\tempurl}


\bibitem[Birhane and Cummins(2019)]%
        {Birhane2019AlgorithmicIT}
\bibfield{author}{\bibinfo{person}{Abeba Birhane} {and} \bibinfo{person}{Fred
  Cummins}.} \bibinfo{year}{2019}\natexlab{}.
\newblock \showarticletitle{Algorithmic Injustices: Towards a Relational
  Ethics}.
\newblock \bibinfo{journal}{\emph{ArXiv}}  \bibinfo{volume}{abs/1912.07376}
  (\bibinfo{year}{2019}).
\newblock


\bibitem[Birhane and Guest(2020)]%
        {birhane2020towards}
\bibfield{author}{\bibinfo{person}{Abeba Birhane} {and} \bibinfo{person}{Olivia
  Guest}.} \bibinfo{year}{2020}\natexlab{}.
\newblock \showarticletitle{Towards decolonising computational sciences}.
\newblock \bibinfo{journal}{\emph{arXiv preprint arXiv:2009.14258}}
  (\bibinfo{year}{2020}).
\newblock


\bibitem[Birhane et~al\mbox{.}(2022)]%
        {Birhane2022FM}
\bibfield{author}{\bibinfo{person}{Abeba Birhane}, \bibinfo{person}{Elayne
  Ruane}, \bibinfo{person}{Thomas Laurent}, \bibinfo{person}{Matthew S.~Brown},
  \bibinfo{person}{Johnathan Flowers}, \bibinfo{person}{Anthony Ventresque},
  {and} \bibinfo{person}{Christopher L.~Dancy}.}
  \bibinfo{year}{2022}\natexlab{}.
\newblock \showarticletitle{The Forgotten Margins of AI Ethics}. In
  \bibinfo{booktitle}{\emph{2022 ACM Conference on Fairness, Accountability,
  and Transparency}} (Seoul, Republic of Korea) \emph{(\bibinfo{series}{FAccT
  '22})}. \bibinfo{publisher}{Association for Computing Machinery},
  \bibinfo{address}{New York, NY, USA}, \bibinfo{pages}{948–958}.
\newblock
\showISBNx{9781450393522}
\urldef\tempurl%
\url{https://doi.org/10.1145/3531146.3533157}
\showDOI{\tempurl}


\bibitem[Blili-Hamelin and Hancox-Li(2022)]%
        {BliliHamelin2022MakingIE}
\bibfield{author}{\bibinfo{person}{Borhane Blili-Hamelin} {and}
  \bibinfo{person}{Leif Hancox-Li}.} \bibinfo{year}{2022}\natexlab{}.
\newblock \showarticletitle{Making Intelligence: Ethical Values in IQ and ML
  Benchmarks}.
\newblock


\bibitem[Bowleg(2021)]%
        {Bowleg2021EvolvingIW}
\bibfield{author}{\bibinfo{person}{Lisa Bowleg}.}
  \bibinfo{year}{2021}\natexlab{}.
\newblock \showarticletitle{Evolving Intersectionality Within Public Health:
  From Analysis to Action.}
\newblock \bibinfo{journal}{\emph{American journal of public health}}
  \bibinfo{volume}{111 1} (\bibinfo{year}{2021}), \bibinfo{pages}{88--90}.
\newblock


\bibitem[Brock(2018)]%
        {Brock2018CriticalTD}
\bibfield{author}{\bibinfo{person}{Andr{\'e} Brock}.}
  \bibinfo{year}{2018}\natexlab{}.
\newblock \showarticletitle{Critical technocultural discourse analysis}.
\newblock \bibinfo{journal}{\emph{New Media \& Society}}  \bibinfo{volume}{20}
  (\bibinfo{year}{2018}), \bibinfo{pages}{1012--1030}.
\newblock


\bibitem[Buchanan et~al\mbox{.}(2020)]%
        {Buchanan2020IntersectionalCH}
\bibfield{author}{\bibinfo{person}{Nicole~T. Buchanan},
  \bibinfo{person}{Desdamona Rios}, {and} \bibinfo{person}{Kim~A. Case}.}
  \bibinfo{year}{2020}\natexlab{}.
\newblock \showarticletitle{Intersectional Cultural Humility: Aligning Critical
  Inquiry with Critical Praxis in Psychology}.
\newblock \bibinfo{journal}{\emph{Women \& Therapy}}  \bibinfo{volume}{43}
  (\bibinfo{year}{2020}), \bibinfo{pages}{235--243}.
\newblock


\bibitem[Buolamwini and Gebru(2018)]%
        {P13}
\bibfield{author}{\bibinfo{person}{Joy Buolamwini} {and}
  \bibinfo{person}{Timnit Gebru}.} \bibinfo{year}{2018}\natexlab{}.
\newblock \showarticletitle{Gender Shades: Intersectional Accuracy Disparities
  in Commercial Gender Classification}. In \bibinfo{booktitle}{\emph{FAT}}.
\newblock


\bibitem[Cabrera et~al\mbox{.}(2019)]%
        {P20}
\bibfield{author}{\bibinfo{person}{{\'A}ngel~Alexander Cabrera},
  \bibinfo{person}{Will Epperson}, \bibinfo{person}{Fred Hohman},
  \bibinfo{person}{Minsuk Kahng}, \bibinfo{person}{Jamie~H. Morgenstern}, {and}
  \bibinfo{person}{Duen~Horng Chau}.} \bibinfo{year}{2019}\natexlab{}.
\newblock \showarticletitle{FAIRVIS: Visual Analytics for Discovering
  Intersectional Bias in Machine Learning}.
\newblock \bibinfo{journal}{\emph{2019 IEEE Conference on Visual Analytics
  Science and Technology (VAST)}} (\bibinfo{year}{2019}),
  \bibinfo{pages}{46--56}.
\newblock


\bibitem[Camara et~al\mbox{.}(2022)]%
        {P24}
\bibfield{author}{\bibinfo{person}{Antonio Camara}, \bibinfo{person}{Nina
  Taneja}, \bibinfo{person}{Tamjeed Azad}, \bibinfo{person}{Emily Allaway},
  {and} \bibinfo{person}{Richard~S. Zemel}.} \bibinfo{year}{2022}\natexlab{}.
\newblock \showarticletitle{Mapping the Multilingual Margins: Intersectional
  Biases of Sentiment Analysis Systems in English, Spanish, and Arabic}. In
  \bibinfo{booktitle}{\emph{LTEDI}}.
\newblock


\bibitem[Cameron(2004)]%
        {cameron2004evidence}
\bibfield{author}{\bibinfo{person}{Graham Cameron}.}
  \bibinfo{year}{2004}\natexlab{}.
\newblock \showarticletitle{Evidence in an indigenous world}. In
  \bibinfo{booktitle}{\emph{Australasian Evaluation Society 2004 International
  Conference, Adelaide, South Australia}}.
\newblock


\bibitem[Chilisa(2019)]%
        {chilisa2019indigenous}
\bibfield{author}{\bibinfo{person}{Bagele Chilisa}.}
  \bibinfo{year}{2019}\natexlab{}.
\newblock \bibinfo{booktitle}{\emph{Indigenous research methodologies}}.
\newblock \bibinfo{publisher}{Sage publications}.
\newblock


\bibitem[Cho et~al\mbox{.}(2013)]%
        {Cho_Crenshaw_McCall_2013}
\bibfield{author}{\bibinfo{person}{Sumi Cho},
  \bibinfo{person}{Kimberlé~Williams Crenshaw}, {and} \bibinfo{person}{Leslie
  McCall}.} \bibinfo{year}{2013}\natexlab{}.
\newblock \showarticletitle{Toward a Field of Intersectionality Studies:
  Theory, Applications, and Praxis}.
\newblock \bibinfo{journal}{\emph{Signs: Journal of Women in Culture and
  Society}} \bibinfo{volume}{38}, \bibinfo{number}{4} (\bibinfo{date}{Jun}
  \bibinfo{year}{2013}), \bibinfo{pages}{785–810}.
\newblock
\showISSN{0097-9740, 1545-6943}
\urldef\tempurl%
\url{https://doi.org/10.1086/669608}
\showDOI{\tempurl}


\bibitem[Ciston(2019)]%
        {ciston2019imagining}
\bibfield{author}{\bibinfo{person}{Sarah Ciston}.}
  \bibinfo{year}{2019}\natexlab{}.
\newblock \showarticletitle{Imagining Intersectional AI}.
\newblock \bibinfo{journal}{\emph{xCoAx}} (\bibinfo{year}{2019}),
  \bibinfo{pages}{39}.
\newblock


\bibitem[Cole(2009)]%
        {Cole2009IntersectionalityAR}
\bibfield{author}{\bibinfo{person}{Elizabeth~R Cole}.}
  \bibinfo{year}{2009}\natexlab{}.
\newblock \showarticletitle{Intersectionality and research in psychology.}
\newblock \bibinfo{journal}{\emph{The American psychologist}}
  \bibinfo{volume}{64 3} (\bibinfo{year}{2009}), \bibinfo{pages}{170--80}.
\newblock


\bibitem[Collective(1978)]%
        {The_Combahee_River_Collective_1978}
\bibfield{author}{\bibinfo{person}{The Combahee~River Collective}.}
  \bibinfo{year}{1978}\natexlab{}.
\newblock \showarticletitle{A Black Feminist Statement}.
\newblock \bibinfo{journal}{\emph{Women’s Studies Quarterly}}
  (\bibinfo{year}{1978}).
\newblock


\bibitem[Collins(2000)]%
        {Collins2017BlackFT}
\bibfield{author}{\bibinfo{person}{Patricia~Hill Collins}.}
  \bibinfo{year}{2000}\natexlab{}.
\newblock \showarticletitle{Black Feminist Thought in the Matrix of
  Domination}.
\newblock


\bibitem[Collins(2015)]%
        {Collins2015IntersectionalitysDD}
\bibfield{author}{\bibinfo{person}{Patr{\'i}cia~Hill Collins}.}
  \bibinfo{year}{2015}\natexlab{}.
\newblock \showarticletitle{Intersectionality's Definitional Dilemmas}.
\newblock \bibinfo{journal}{\emph{Review of Sociology}}  \bibinfo{volume}{41}
  (\bibinfo{year}{2015}), \bibinfo{pages}{1--20}.
\newblock


\bibitem[Collins(2019)]%
        {collins2019intersectionality}
\bibfield{author}{\bibinfo{person}{Patricia~Hill Collins}.}
  \bibinfo{year}{2019}\natexlab{}.
\newblock \bibinfo{booktitle}{\emph{Intersectionality as critical social
  theory}}.
\newblock \bibinfo{publisher}{Duke University Press}.
\newblock


\bibitem[Collins and Bilge(2020)]%
        {collins2020intersectionality}
\bibfield{author}{\bibinfo{person}{Patricia~Hill Collins} {and}
  \bibinfo{person}{Sirma Bilge}.} \bibinfo{year}{2020}\natexlab{}.
\newblock \bibinfo{booktitle}{\emph{Intersectionality}}.
\newblock \bibinfo{publisher}{John Wiley \& Sons}.
\newblock


\bibitem[Collins et~al\mbox{.}(2019)]%
        {Collins2019IntersectionalityAC}
\bibfield{author}{\bibinfo{person}{Patricia~Hill Collins},
  \bibinfo{person}{Elaini Cristina~Gonzaga da Silva}, \bibinfo{person}{Emek
  Ergun}, \bibinfo{person}{Inger Furseth}, \bibinfo{person}{Kanisha~D. Bond},
  {and} \bibinfo{person}{Jone Mart{\'i}nez-Palacios}.}
  \bibinfo{year}{2019}\natexlab{}.
\newblock \showarticletitle{Intersectionality as Critical Social Theory}.
\newblock \bibinfo{journal}{\emph{Contemporary Political Theory}}
  \bibinfo{volume}{20} (\bibinfo{year}{2019}), \bibinfo{pages}{690--725}.
\newblock


\bibitem[Constanza-Chock(2020)]%
        {2020IntroductionD}
\bibfield{author}{\bibinfo{person}{Sasha Constanza-Chock}.}
  \bibinfo{year}{2020}\natexlab{}.
\newblock \showarticletitle{Introduction: \#TravelingWhileTrans, Design
  Justice, and Escape from the Matrix of Domination}.
\newblock \bibinfo{journal}{\emph{Design Justice}} (\bibinfo{year}{2020}).
\newblock


\bibitem[Cooper et~al\mbox{.}(2021)]%
        {Cooper2021EmergentUI}
\bibfield{author}{\bibinfo{person}{A.~Feder Cooper}, \bibinfo{person}{Ellen
  Abrams}, {and} \bibinfo{person}{NA Na}.} \bibinfo{year}{2021}\natexlab{}.
\newblock \showarticletitle{Emergent Unfairness in Algorithmic
  Fairness-Accuracy Trade-Off Research}.
\newblock \bibinfo{journal}{\emph{Proceedings of the 2021 AAAI/ACM Conference
  on AI, Ethics, and Society}} (\bibinfo{year}{2021}).
\newblock


\bibitem[Cowan(1972)]%
        {cowan1972francis}
\bibfield{author}{\bibinfo{person}{Ruth~Schwartz Cowan}.}
  \bibinfo{year}{1972}\natexlab{}.
\newblock \showarticletitle{Francis Galton's statistical ideas: the influence
  of eugenics}.
\newblock \bibinfo{journal}{\emph{Isis}} \bibinfo{volume}{63},
  \bibinfo{number}{4} (\bibinfo{year}{1972}), \bibinfo{pages}{509--528}.
\newblock


\bibitem[Cram(2004)]%
        {cram2004kaupapa}
\bibfield{author}{\bibinfo{person}{Fiona Cram}.}
  \bibinfo{year}{2004}\natexlab{}.
\newblock \showarticletitle{Kaupapa M{\=a}ori evaluation: Theories, practices,
  models, analyses}.
\newblock \bibinfo{journal}{\emph{Evaluation Hui Summit}}
  (\bibinfo{year}{2004}), \bibinfo{pages}{11--16}.
\newblock


\bibitem[Crenshaw(1989)]%
        {crenshaw1989demarginalizing}
\bibfield{author}{\bibinfo{person}{Kimberl{\'e} Crenshaw}.}
  \bibinfo{year}{1989}\natexlab{}.
\newblock \showarticletitle{Demarginalizing the intersection of race and sex: A
  black feminist critique of antidiscrimination doctrine, feminist theory and
  antiracist politics}.
\newblock In \bibinfo{booktitle}{\emph{Feminist legal theories}}.
  \bibinfo{publisher}{Routledge}, \bibinfo{pages}{23--51}.
\newblock


\bibitem[Crenshaw(1991)]%
        {Crenshaw_1991}
\bibfield{author}{\bibinfo{person}{Kimberle Crenshaw}.}
  \bibinfo{year}{1991}\natexlab{}.
\newblock \showarticletitle{Mapping the Margins: Intersectionality, Identity
  Politics, and Violence against Women of Color}.
\newblock \bibinfo{journal}{\emph{Stanford Law Review}} \bibinfo{volume}{43},
  \bibinfo{number}{6} (\bibinfo{date}{Jul} \bibinfo{year}{1991}),
  \bibinfo{pages}{1241}.
\newblock
\showISSN{00389765}
\urldef\tempurl%
\url{https://doi.org/10.2307/1229039}
\showDOI{\tempurl}


\bibitem[Davis et~al\mbox{.}(2021)]%
        {P16}
\bibfield{author}{\bibinfo{person}{Jenny~L. Davis}, \bibinfo{person}{Apryl~A.
  Williams}, {and} \bibinfo{person}{Michael~W. Yang}.}
  \bibinfo{year}{2021}\natexlab{}.
\newblock \showarticletitle{Algorithmic reparation}.
\newblock \bibinfo{journal}{\emph{Big Data \& Society}}  \bibinfo{volume}{8}
  (\bibinfo{year}{2021}).
\newblock


\bibitem[Delgado and Stefancic(2023)]%
        {delgado2023critical}
\bibfield{author}{\bibinfo{person}{Richard Delgado} {and} \bibinfo{person}{Jean
  Stefancic}.} \bibinfo{year}{2023}\natexlab{}.
\newblock \bibinfo{booktitle}{\emph{Critical race theory: An introduction}}.
  Vol.~\bibinfo{volume}{87}.
\newblock \bibinfo{publisher}{NyU press}.
\newblock


\bibitem[D’Ignazio(2021)]%
        {DIgnazio2021DataFT}
\bibfield{author}{\bibinfo{person}{Catherine D’Ignazio}.}
  \bibinfo{year}{2021}\natexlab{}.
\newblock \showarticletitle{Data Feminism: Teaching and Learning for Justice}.
\newblock \bibinfo{journal}{\emph{Proceedings of the 26th ACM Conference on
  Innovation and Technology in Computer Science Education V. 1}}
  (\bibinfo{year}{2021}).
\newblock


\bibitem[Eubanks(2018)]%
        {Eubanks2018AutomatingIH}
\bibfield{author}{\bibinfo{person}{Virginia~E. Eubanks}.}
  \bibinfo{year}{2018}\natexlab{}.
\newblock \showarticletitle{Automating Inequality: How High-Tech Tools Profile,
  Police, and Punish the Poor}.
\newblock


\bibitem[Fanon(1952)]%
        {Fanon1952BlackSW}
\bibfield{author}{\bibinfo{person}{Frantz Fanon}.}
  \bibinfo{year}{1952}\natexlab{}.
\newblock \showarticletitle{Black Skin, White Masks}.
\newblock \bibinfo{journal}{\emph{My Black Stars}} (\bibinfo{year}{1952}).
\newblock


\bibitem[Fanon(2004)]%
        {fanon2004wretched}
\bibfield{author}{\bibinfo{person}{Frantz Fanon}.}
  \bibinfo{year}{2004}\natexlab{}.
\newblock \showarticletitle{The Wretched of the Earth. 1961}.
\newblock \bibinfo{journal}{\emph{Trans. Richard Philcox. New York: Grove
  Press}}  \bibinfo{volume}{6} (\bibinfo{year}{2004}).
\newblock


\bibitem[Finocchiaro et~al\mbox{.}(2020)]%
        {P29}
\bibfield{author}{\bibinfo{person}{Jessica Finocchiaro},
  \bibinfo{person}{Roland Maio}, \bibinfo{person}{Faidra~Georgia Monachou},
  \bibinfo{person}{Gourab~K. Patro}, \bibinfo{person}{Manish Raghavan},
  \bibinfo{person}{Ana-Andreea Stoica}, {and} \bibinfo{person}{Stratis
  Tsirtsis}.} \bibinfo{year}{2020}\natexlab{}.
\newblock \showarticletitle{Bridging Machine Learning and Mechanism Design
  towards Algorithmic Fairness}.
\newblock \bibinfo{journal}{\emph{Proceedings of the 2021 ACM Conference on
  Fairness, Accountability, and Transparency}} (\bibinfo{year}{2020}).
\newblock


\bibitem[Fitzsimons et~al\mbox{.}(2018)]%
        {P14}
\bibfield{author}{\bibinfo{person}{Jack~K. Fitzsimons},
  \bibinfo{person}{Michael~A. Osborne}, {and} \bibinfo{person}{Stephen~J.
  Roberts}.} \bibinfo{year}{2018}\natexlab{}.
\newblock \showarticletitle{Intersectionality: Multiple Group Fairness in
  Expectation Constraints}.
\newblock \bibinfo{journal}{\emph{ArXiv}}  \bibinfo{volume}{abs/1811.09960}
  (\bibinfo{year}{2018}).
\newblock


\bibitem[Foulds et~al\mbox{.}(2018)]%
        {P2}
\bibfield{author}{\bibinfo{person}{James~R. Foulds}, \bibinfo{person}{Rashidul
  Islam}, \bibinfo{person}{Kamrun Keya}, {and} \bibinfo{person}{Shimei Pan}.}
  \bibinfo{year}{2018}\natexlab{}.
\newblock \showarticletitle{Bayesian Modeling of Intersectional Fairness: The
  Variance of Bias}. In \bibinfo{booktitle}{\emph{SDM}}.
\newblock


\bibitem[Foulds and Pan(2018)]%
        {P5}
\bibfield{author}{\bibinfo{person}{James~R. Foulds} {and}
  \bibinfo{person}{Shimei Pan}.} \bibinfo{year}{2018}\natexlab{}.
\newblock \showarticletitle{An Intersectional Definition of Fairness}.
\newblock \bibinfo{journal}{\emph{2020 IEEE 36th International Conference on
  Data Engineering (ICDE)}} (\bibinfo{year}{2018}),
  \bibinfo{pages}{1918--1921}.
\newblock


\bibitem[Freeman(1977)]%
        {freeman1977legitimizing}
\bibfield{author}{\bibinfo{person}{Alan~David Freeman}.}
  \bibinfo{year}{1977}\natexlab{}.
\newblock \showarticletitle{Legitimizing racial discrimination through
  antidiscrimination law: A critical review of Supreme Court doctrine}.
\newblock \bibinfo{journal}{\emph{Minn. L. Rev.}}  \bibinfo{volume}{62}
  (\bibinfo{year}{1977}), \bibinfo{pages}{1049}.
\newblock


\bibitem[Gebru(2021)]%
        {Gebru_2021}
\bibfield{author}{\bibinfo{person}{Timnit Gebru}.}
  \bibinfo{year}{2021}\natexlab{}.
\newblock \bibinfo{title}{Hierarchy of Knowledge in Machine Learning \& Related
  Fields and Its Consequence}.
\newblock
\newblock
\urldef\tempurl%
\url{https://youtu.be/OL3DowBM9uc}
\showURL{%
\tempurl}


\bibitem[Ghosh et~al\mbox{.}(2021)]%
        {P15}
\bibfield{author}{\bibinfo{person}{A. Ghosh}, \bibinfo{person}{Lea Genuit},
  {and} \bibinfo{person}{Mary Reagan}.} \bibinfo{year}{2021}\natexlab{}.
\newblock \showarticletitle{Characterizing Intersectional Group Fairness with
  Worst-Case Comparisons}. In \bibinfo{booktitle}{\emph{AIDBEI}}.
\newblock


\bibitem[Gordon(2006)]%
        {Gordon_2006}
\bibfield{author}{\bibinfo{person}{Lewis~R. Gordon}.}
  \bibinfo{year}{2006}\natexlab{}.
\newblock \bibinfo{booktitle}{\emph{Is the Human a Teleological Suspension of
  Man? Phenomenological Exploration of Sylvia Wynter’s Fanonian and Biodicean
  Reflections}}.
\newblock \bibinfo{publisher}{Ian Randle}, \bibinfo{address}{Kingston,
  Jamaica}.
\newblock
\showISBNx{978-976-637-741-0}


\bibitem[Guidroz and Berger(2021)]%
        {Guidroz_Berger}
\bibfield{author}{\bibinfo{person}{Kathleen Guidroz} {and}
  \bibinfo{person}{Michele~Tracy Berger}.} \bibinfo{year}{2021}\natexlab{}.
\newblock \showarticletitle{A Conversation with Founding Scholars of
  Intersectionality}.
\newblock  (\bibinfo{year}{2021}).
\newblock


\bibitem[Hampton(2021)]%
        {Hampton2021BlackFM}
\bibfield{author}{\bibinfo{person}{Lelia Hampton}.}
  \bibinfo{year}{2021}\natexlab{}.
\newblock \showarticletitle{Black Feminist Musings on Algorithmic Oppression}.
\newblock \bibinfo{journal}{\emph{Proceedings of the 2021 ACM Conference on
  Fairness, Accountability, and Transparency}} (\bibinfo{year}{2021}).
\newblock


\bibitem[Hancock(2007)]%
        {Hancock_2007}
\bibfield{author}{\bibinfo{person}{Ange-Marie Hancock}.}
  \bibinfo{year}{2007}\natexlab{}.
\newblock \showarticletitle{When Multiplication Doesn’t Equal Quick Addition:
  Examining Intersectionality as a Research Paradigm}.
\newblock \bibinfo{journal}{\emph{Perspectives on Politics}}
  \bibinfo{volume}{5}, \bibinfo{number}{01} (\bibinfo{date}{Mar}
  \bibinfo{year}{2007}).
\newblock
\showISSN{1537-5927, 1541-0986}
\urldef\tempurl%
\url{https://doi.org/10.1017/S1537592707070065}
\showDOI{\tempurl}


\bibitem[Jacobs and Wallach(2019)]%
        {Jacobs2019MeasurementAF}
\bibfield{author}{\bibinfo{person}{Abigail~Z. Jacobs} {and}
  \bibinfo{person}{Hanna~M. Wallach}.} \bibinfo{year}{2019}\natexlab{}.
\newblock \showarticletitle{Measurement and Fairness}.
\newblock \bibinfo{journal}{\emph{Proceedings of the 2021 ACM Conference on
  Fairness, Accountability, and Transparency}} (\bibinfo{year}{2019}).
\newblock


\bibitem[Jin et~al\mbox{.}(2020)]%
        {P22}
\bibfield{author}{\bibinfo{person}{Zhongjun~(Mark) Jin},
  \bibinfo{person}{Mengjing Xu}, \bibinfo{person}{Chenkai Sun},
  \bibinfo{person}{Abolfazl Asudeh}, {and} \bibinfo{person}{H.~V. Jagadish}.}
  \bibinfo{year}{2020}\natexlab{}.
\newblock \showarticletitle{MithraCoverage: A System for Investigating
  Population Bias for Intersectional Fairness}.
\newblock \bibinfo{journal}{\emph{Proceedings of the 2020 ACM SIGMOD
  International Conference on Management of Data}} (\bibinfo{year}{2020}).
\newblock


\bibitem[Kalichman et~al\mbox{.}(2021)]%
        {Kalichman2021FindingTR}
\bibfield{author}{\bibinfo{person}{Seth~C. Kalichman}, \bibinfo{person}{Bruno
  Shkembi}, {and} \bibinfo{person}{Lisa~A. Eaton}.}
  \bibinfo{year}{2021}\natexlab{}.
\newblock \showarticletitle{Finding the Right Angle: A Geometric Approach to
  Measuring Intersectional HIV Stigma}.
\newblock \bibinfo{journal}{\emph{AIDS and Behavior}}  \bibinfo{volume}{26}
  (\bibinfo{year}{2021}), \bibinfo{pages}{27--38}.
\newblock


\bibitem[Kang et~al\mbox{.}(2021)]%
        {P21}
\bibfield{author}{\bibinfo{person}{Jian Kang}, \bibinfo{person}{Tiankai Xie},
  \bibinfo{person}{Xintao Wu}, \bibinfo{person}{Ross Maciejewski}, {and}
  \bibinfo{person}{Hanghang Tong}.} \bibinfo{year}{2021}\natexlab{}.
\newblock \showarticletitle{InfoFair: Information-Theoretic Intersectional
  Fairness}.
\newblock \bibinfo{journal}{\emph{2022 IEEE International Conference on Big
  Data (Big Data)}} (\bibinfo{year}{2021}), \bibinfo{pages}{1455--1464}.
\newblock


\bibitem[Kannan et~al\mbox{.}(2019)]%
        {kannan2019downstream}
\bibfield{author}{\bibinfo{person}{Sampath Kannan}, \bibinfo{person}{Aaron
  Roth}, {and} \bibinfo{person}{Juba Ziani}.} \bibinfo{year}{2019}\natexlab{}.
\newblock \showarticletitle{Downstream effects of affirmative action}. In
  \bibinfo{booktitle}{\emph{Proceedings of the Conference on Fairness,
  Accountability, and Transparency}}. \bibinfo{pages}{240--248}.
\newblock


\bibitem[Kasy and Abebe(2021)]%
        {P19}
\bibfield{author}{\bibinfo{person}{Maximilian Kasy} {and}
  \bibinfo{person}{Rediet Abebe}.} \bibinfo{year}{2021}\natexlab{}.
\newblock \showarticletitle{Fairness, Equality, and Power in Algorithmic
  Decision-Making}.
\newblock \bibinfo{journal}{\emph{Proceedings of the 2021 ACM Conference on
  Fairness, Accountability, and Transparency}} (\bibinfo{year}{2021}).
\newblock


\bibitem[Kendi(2023)]%
        {Kendi_2023}
\bibfield{author}{\bibinfo{person}{Ibram~X. Kendi}.}
  \bibinfo{year}{2023}\natexlab{}.
\newblock \bibinfo{booktitle}{\emph{How to be an antiracist}}.
\newblock \bibinfo{publisher}{One World}.
\newblock


\bibitem[Kim et~al\mbox{.}(2020)]%
        {P11}
\bibfield{author}{\bibinfo{person}{Jae-Yeon Kim}, \bibinfo{person}{Carlos
  Ortiz}, \bibinfo{person}{Sarah Nam}, \bibinfo{person}{Sarah Santiago}, {and}
  \bibinfo{person}{Vivek Datta}.} \bibinfo{year}{2020}\natexlab{}.
\newblock \showarticletitle{Intersectional Bias in Hate Speech and Abusive
  Language Datasets}.
\newblock \bibinfo{journal}{\emph{ArXiv}}  \bibinfo{volume}{abs/2005.05921}
  (\bibinfo{year}{2020}).
\newblock


\bibitem[Kirk et~al\mbox{.}(2021)]%
        {P10}
\bibfield{author}{\bibinfo{person}{Hannah~Rose Kirk}, \bibinfo{person}{Yennie
  Jun}, \bibinfo{person}{Haider Iqbal}, \bibinfo{person}{Elias Benussi},
  \bibinfo{person}{Filippo Volpin}, \bibinfo{person}{Fr{\'e}d{\'e}ric~A.
  Dreyer}, \bibinfo{person}{Aleksandar Shtedritski}, {and}
  \bibinfo{person}{Yuki~M. Asano}.} \bibinfo{year}{2021}\natexlab{}.
\newblock \showarticletitle{Bias Out-of-the-Box: An Empirical Analysis of
  Intersectional Occupational Biases in Popular Generative Language Models}. In
  \bibinfo{booktitle}{\emph{Neural Information Processing Systems}}.
\newblock


\bibitem[Klumbyte et~al\mbox{.}(2022a)]%
        {klumbyte2022critical}
\bibfield{author}{\bibinfo{person}{Goda Klumbyte}, \bibinfo{person}{Claude
  Draude}, {and} \bibinfo{person}{Alex~S. Taylor}.}
  \bibinfo{year}{2022}\natexlab{a}.
\newblock \showarticletitle{Critical Tools for Machine Learning: Working with
  Intersectional Critical Concepts in Machine Learning Systems Design}. In
  \bibinfo{booktitle}{\emph{2022 ACM Conference on Fairness, Accountability,
  and Transparency}} (Seoul, Republic of Korea) \emph{(\bibinfo{series}{FAccT
  '22})}. \bibinfo{publisher}{Association for Computing Machinery},
  \bibinfo{address}{New York, NY, USA}, \bibinfo{pages}{1528–1541}.
\newblock
\showISBNx{9781450393522}
\urldef\tempurl%
\url{https://doi.org/10.1145/3531146.3533207}
\showDOI{\tempurl}


\bibitem[Klumbyte et~al\mbox{.}(2022b)]%
        {P8}
\bibfield{author}{\bibinfo{person}{Goda Klumbyte}, \bibinfo{person}{Claude
  Draude}, {and} \bibinfo{person}{Alex~S. Taylor}.}
  \bibinfo{year}{2022}\natexlab{b}.
\newblock \showarticletitle{Critical Tools for Machine Learning: Working with
  Intersectional Critical Concepts in Machine Learning Systems Design}.
\newblock \bibinfo{journal}{\emph{2022 ACM Conference on Fairness,
  Accountability, and Transparency}} (\bibinfo{year}{2022}).
\newblock


\bibitem[Kobayashi and Nakao(2020)]%
        {P9}
\bibfield{author}{\bibinfo{person}{Kenji Kobayashi} {and} \bibinfo{person}{Yuri
  Nakao}.} \bibinfo{year}{2020}\natexlab{}.
\newblock \showarticletitle{One-vs.-One Mitigation of Intersectional Bias: A
  General Method to Extend Fairness-Aware Binary Classification}.
\newblock \bibinfo{journal}{\emph{ArXiv}}  \bibinfo{volume}{abs/2010.13494}
  (\bibinfo{year}{2020}).
\newblock


\bibitem[Kong(2022a)]%
        {P3}
\bibfield{author}{\bibinfo{person}{Youjin Kong}.}
  \bibinfo{year}{2022}\natexlab{a}.
\newblock \showarticletitle{Are “Intersectionally Fair” AI Algorithms
  Really Fair to Women of Color? A Philosophical Analysis}.
\newblock \bibinfo{journal}{\emph{2022 ACM Conference on Fairness,
  Accountability, and Transparency}} (\bibinfo{year}{2022}).
\newblock


\bibitem[Kong(2022b)]%
        {kong2022intersectionally}
\bibfield{author}{\bibinfo{person}{Youjin Kong}.}
  \bibinfo{year}{2022}\natexlab{b}.
\newblock \showarticletitle{Are “Intersectionally Fair” AI Algorithms
  Really Fair to Women of Color? A Philosophical Analysis}. In
  \bibinfo{booktitle}{\emph{2022 ACM Conference on Fairness, Accountability,
  and Transparency}}. \bibinfo{pages}{485--494}.
\newblock


\bibitem[Lalor et~al\mbox{.}(2022)]%
        {P4}
\bibfield{author}{\bibinfo{person}{John~P. Lalor}, \bibinfo{person}{Yi Yang},
  \bibinfo{person}{Kendall Smith}, \bibinfo{person}{Nicole Forsgren}, {and}
  \bibinfo{person}{A. Abbasi}.} \bibinfo{year}{2022}\natexlab{}.
\newblock \showarticletitle{Benchmarking Intersectional Biases in NLP}. In
  \bibinfo{booktitle}{\emph{North American Chapter of the Association for
  Computational Linguistics}}.
\newblock


\bibitem[Makhlouf et~al\mbox{.}(2021)]%
        {P30}
\bibfield{author}{\bibinfo{person}{Karima Makhlouf}, \bibinfo{person}{Sami
  Zhioua}, {and} \bibinfo{person}{Catuscia Palamidessi}.}
  \bibinfo{year}{2021}\natexlab{}.
\newblock \showarticletitle{On the Applicability of Machine Learning Fairness
  Notions}.
\newblock \bibinfo{journal}{\emph{ACM SIGKDD Explorations Newsletter}}
  \bibinfo{volume}{23} (\bibinfo{year}{2021}), \bibinfo{pages}{14--23}.
\newblock


\bibitem[Mhasawade et~al\mbox{.}(2021)]%
        {P23}
\bibfield{author}{\bibinfo{person}{Vishwali Mhasawade}, \bibinfo{person}{Yuan
  Zhao}, {and} \bibinfo{person}{Rumi Chunara}.}
  \bibinfo{year}{2021}\natexlab{}.
\newblock \showarticletitle{Machine learning and algorithmic fairness in public
  and population health}.
\newblock \bibinfo{journal}{\emph{Nature Machine Intelligence}}
  \bibinfo{volume}{3} (\bibinfo{year}{2021}), \bibinfo{pages}{659--666}.
\newblock


\bibitem[Mitchell et~al\mbox{.}(2018)]%
        {P18}
\bibfield{author}{\bibinfo{person}{Margaret Mitchell}, \bibinfo{person}{Simone
  Wu}, \bibinfo{person}{Andrew Zaldivar}, \bibinfo{person}{Parker Barnes},
  \bibinfo{person}{Lucy Vasserman}, \bibinfo{person}{Ben Hutchinson},
  \bibinfo{person}{Elena Spitzer}, \bibinfo{person}{Inioluwa~Deborah Raji},
  {and} \bibinfo{person}{Timnit Gebru}.} \bibinfo{year}{2018}\natexlab{}.
\newblock \showarticletitle{Model Cards for Model Reporting}.
\newblock \bibinfo{journal}{\emph{Proceedings of the Conference on Fairness,
  Accountability, and Transparency}} (\bibinfo{year}{2018}).
\newblock


\bibitem[Mohamed et~al\mbox{.}(2020)]%
        {mohamed2020decolonial}
\bibfield{author}{\bibinfo{person}{Shakir Mohamed},
  \bibinfo{person}{Marie-Therese Png}, {and} \bibinfo{person}{William Isaac}.}
  \bibinfo{year}{2020}\natexlab{}.
\newblock \showarticletitle{Decolonial AI: Decolonial theory as sociotechnical
  foresight in artificial intelligence}.
\newblock \bibinfo{journal}{\emph{Philosophy \& Technology}}
  \bibinfo{volume}{33} (\bibinfo{year}{2020}), \bibinfo{pages}{659--684}.
\newblock


\bibitem[Molina and Loiseau(2022)]%
        {P26}
\bibfield{author}{\bibinfo{person}{Mathieu Molina} {and}
  \bibinfo{person}{Patrick Loiseau}.} \bibinfo{year}{2022}\natexlab{}.
\newblock \showarticletitle{Bounding and Approximating Intersectional Fairness
  through Marginal Fairness}.
\newblock \bibinfo{journal}{\emph{ArXiv}}  \bibinfo{volume}{abs/2206.05828}
  (\bibinfo{year}{2022}).
\newblock


\bibitem[Moug{\'a}n et~al\mbox{.}(2022)]%
        {P28}
\bibfield{author}{\bibinfo{person}{Carlos Moug{\'a}n},
  \bibinfo{person}{Jos{\'e}~Manuel {\'A}lvarez}, \bibinfo{person}{Gourab~K.
  Patro}, \bibinfo{person}{Salvatore Ruggieri}, {and} \bibinfo{person}{Steffen
  Staab}.} \bibinfo{year}{2022}\natexlab{}.
\newblock \showarticletitle{Fairness implications of encoding protected
  categorical attributes}.
\newblock \bibinfo{journal}{\emph{ArXiv}}  \bibinfo{volume}{abs/2201.11358}
  (\bibinfo{year}{2022}).
\newblock


\bibitem[Noble(2018)]%
        {Noble_2018}
\bibfield{author}{\bibinfo{person}{Safiya~Umoja Noble}.}
  \bibinfo{year}{2018}\natexlab{}.
\newblock \bibinfo{booktitle}{\emph{Algorithms of oppression: how search
  engines reinforce racism}}.
\newblock \bibinfo{publisher}{New York University Press}, \bibinfo{address}{New
  York}.
\newblock
\showISBNx{978-1-4798-4994-9}


\bibitem[Nowell et~al\mbox{.}(2017)]%
        {nowell2017thematic}
\bibfield{author}{\bibinfo{person}{Lorelli~S. Nowell}, \bibinfo{person}{Jill~M.
  Norris}, \bibinfo{person}{Deborah~E. White}, {and} \bibinfo{person}{Nancy~J.
  Moules}.} \bibinfo{year}{2017}\natexlab{}.
\newblock \showarticletitle{Thematic Analysis: Striving to Meet the
  Trustworthiness Criteria}.
\newblock \bibinfo{journal}{\emph{International Journal of Qualitative
  Methods}} \bibinfo{volume}{16}, \bibinfo{number}{1} (\bibinfo{year}{2017}),
  \bibinfo{pages}{1609406917733847}.
\newblock
\urldef\tempurl%
\url{https://doi.org/10.1177/1609406917733847}
\showDOI{\tempurl}
\showeprint{https://doi.org/10.1177/1609406917733847}


\bibitem[of~Nottingham(nd)]%
        {nottinghamUnderstandingPragmatic}
\bibfield{author}{\bibinfo{person}{University of Nottingham}.}
  \bibinfo{year}{(n.d.)}\natexlab{}.
\newblock \bibinfo{title}{Understanding Pragmatic Research ---
  nottingham.ac.uk}.
\newblock
  \bibinfo{howpublished}{\url{https://www.nottingham.ac.uk/helmopen/rlos/research-evidence-based-practice/designing-research/types-of-study/understanding-pragmatic-research/section02.html}}.
\newblock
\newblock
\shownote{[Accessed 15-Mar-2023]}.


\bibitem[Palaganas et~al\mbox{.}(2017)]%
        {palaganas2017reflexivity}
\bibfield{author}{\bibinfo{person}{Erlinda~C Palaganas},
  \bibinfo{person}{Marian~C Sanchez}, \bibinfo{person}{Visitacion~P Molintas},
  {and} \bibinfo{person}{Ruel~D Caricativo}.} \bibinfo{year}{2017}\natexlab{}.
\newblock \showarticletitle{Reflexivity in qualitative research: A journey of
  learning.}
\newblock \bibinfo{journal}{\emph{Qualitative Report}} \bibinfo{volume}{22},
  \bibinfo{number}{2} (\bibinfo{year}{2017}).
\newblock


\bibitem[Raji et~al\mbox{.}(2021)]%
        {raji2021exclusionary}
\bibfield{author}{\bibinfo{person}{Inioluwa~Deborah Raji},
  \bibinfo{person}{Morgan~Klaus Scheuerman}, {and} \bibinfo{person}{Razvan
  Amironesei}.} \bibinfo{year}{2021}\natexlab{}.
\newblock \showarticletitle{You Can't Sit With Us: Exclusionary Pedagogy in AI
  Ethics Education}. In \bibinfo{booktitle}{\emph{Proceedings of the 2021 ACM
  Conference on Fairness, Accountability, and Transparency}} (Virtual Event,
  Canada) \emph{(\bibinfo{series}{FAccT '21})}. \bibinfo{publisher}{Association
  for Computing Machinery}, \bibinfo{address}{New York, NY, USA},
  \bibinfo{pages}{515–525}.
\newblock
\showISBNx{9781450383097}
\urldef\tempurl%
\url{https://doi.org/10.1145/3442188.3445914}
\showDOI{\tempurl}


\bibitem[Rankin et~al\mbox{.}(2020)]%
        {Rankin2020IntersectionalityIH}
\bibfield{author}{\bibinfo{person}{Yolanda~A. Rankin},
  \bibinfo{person}{Jakita~Owensby Thomas}, {and} \bibinfo{person}{Nicole~M.
  Joseph}.} \bibinfo{year}{2020}\natexlab{}.
\newblock \showarticletitle{Intersectionality in HCI}.
\newblock \bibinfo{journal}{\emph{Interactions}}  \bibinfo{volume}{27}
  (\bibinfo{year}{2020}), \bibinfo{pages}{68--71}.
\newblock


\bibitem[Rogerson and Fitzsimmons(2022)]%
        {P6}
\bibfield{author}{\bibinfo{person}{Kenneth~S. Rogerson} {and}
  \bibinfo{person}{Aidan Fitzsimmons}.} \bibinfo{year}{2022}\natexlab{}.
\newblock \showarticletitle{Intersectional Identities and Machine Learning:
  Illuminating Language Biases in Twitter Algorithms}.
\newblock \bibinfo{journal}{\emph{Proceedings of the Annual Hawaii
  International Conference on System Sciences}} (\bibinfo{year}{2022}).
\newblock


\bibitem[Schlesinger et~al\mbox{.}(2017)]%
        {schlesinger2017interhci}
\bibfield{author}{\bibinfo{person}{Ari Schlesinger}, \bibinfo{person}{W.~Keith
  Edwards}, {and} \bibinfo{person}{Rebecca~E. Grinter}.}
  \bibinfo{year}{2017}\natexlab{}.
\newblock \showarticletitle{Intersectional HCI: Engaging Identity through
  Gender, Race, and Class}. In \bibinfo{booktitle}{\emph{Proceedings of the
  2017 CHI Conference on Human Factors in Computing Systems}} (Denver,
  Colorado, USA) \emph{(\bibinfo{series}{CHI '17})}.
  \bibinfo{publisher}{Association for Computing Machinery},
  \bibinfo{address}{New York, NY, USA}, \bibinfo{pages}{5412–5427}.
\newblock
\showISBNx{9781450346559}
\urldef\tempurl%
\url{https://doi.org/10.1145/3025453.3025766}
\showDOI{\tempurl}


\bibitem[Smith(2021)]%
        {smith2021decolonizing}
\bibfield{author}{\bibinfo{person}{Linda~Tuhiwai Smith}.}
  \bibinfo{year}{2021}\natexlab{}.
\newblock \bibinfo{booktitle}{\emph{Decolonizing methodologies: Research and
  indigenous peoples}}.
\newblock \bibinfo{publisher}{Bloomsbury Publishing}.
\newblock


\bibitem[Spade(2015)]%
        {spade2015normal}
\bibfield{author}{\bibinfo{person}{Dean Spade}.}
  \bibinfo{year}{2015}\natexlab{}.
\newblock \bibinfo{booktitle}{\emph{Normal life: Administrative violence,
  critical trans politics, and the limits of law}}.
\newblock \bibinfo{publisher}{Duke University Press}.
\newblock


\bibitem[Steed and Caliskan(2020)]%
        {P17}
\bibfield{author}{\bibinfo{person}{Ryan Steed} {and} \bibinfo{person}{Aylin
  Caliskan}.} \bibinfo{year}{2020}\natexlab{}.
\newblock \showarticletitle{Image Representations Learned With Unsupervised
  Pre-Training Contain Human-like Biases}.
\newblock \bibinfo{journal}{\emph{Proceedings of the 2021 ACM Conference on
  Fairness, Accountability, and Transparency}} (\bibinfo{year}{2020}).
\newblock


\bibitem[Stent(2023)]%
        {howtorea21:online}
\bibfield{author}{\bibinfo{person}{Amanda Stent}.}
  \bibinfo{year}{2023}\natexlab{}.
\newblock \bibinfo{title}{howtoreadacspaper.pdf}.
\newblock
  \bibinfo{howpublished}{\url{https://people.cs.pitt.edu/~litman/courses/cs2710/papers/howtoreadacspaper.pdf}}.
\newblock
\newblock
\shownote{(Accessed on 03/06/2023)}.


\bibitem[Suresh et~al\mbox{.}(2022)]%
        {P7}
\bibfield{author}{\bibinfo{person}{Harini Suresh}, \bibinfo{person}{Rajiv
  Movva}, \bibinfo{person}{Amelia~Lee Dogan}, \bibinfo{person}{Rahul Bhargava},
  \bibinfo{person}{Isadora~Araujo Crux{\^e}n},
  \bibinfo{person}{Angeles~Martinez Cuba}, \bibinfo{person}{Guilia Taurino},
  \bibinfo{person}{Wonyoung So}, {and} \bibinfo{person}{Catherine
  D’Ignazio}.} \bibinfo{year}{2022}\natexlab{}.
\newblock \showarticletitle{Towards Intersectional Feminist and Participatory
  ML: A Case Study in Supporting Feminicide Counterdata Collection}.
\newblock \bibinfo{journal}{\emph{2022 ACM Conference on Fairness,
  Accountability, and Transparency}} (\bibinfo{year}{2022}).
\newblock


\bibitem[Sweeney and Brock(2014)]%
        {sweeney2014critical}
\bibfield{author}{\bibinfo{person}{Miriam~E Sweeney} {and}
  \bibinfo{person}{Andr{\'e} Brock}.} \bibinfo{year}{2014}\natexlab{}.
\newblock \showarticletitle{Critical informatics: New methods and practices}.
\newblock \bibinfo{journal}{\emph{Proceedings of the American Society for
  Information Science and Technology}} \bibinfo{volume}{51},
  \bibinfo{number}{1} (\bibinfo{year}{2014}), \bibinfo{pages}{1--8}.
\newblock


\bibitem[Talat et~al\mbox{.}(2021)]%
        {Talat2021DisembodiedML}
\bibfield{author}{\bibinfo{person}{Zeerak Talat}, \bibinfo{person}{Joachim
  Bingel}, {and} \bibinfo{person}{Isabelle Augenstein}.}
  \bibinfo{year}{2021}\natexlab{}.
\newblock \showarticletitle{Disembodied Machine Learning: On the Illusion of
  Objectivity in NLP}.
\newblock \bibinfo{journal}{\emph{ArXiv}}  \bibinfo{volume}{abs/2101.11974}
  (\bibinfo{year}{2021}).
\newblock


\bibitem[Thambinathan and Kinsella(2021)]%
        {thambinathan2021decolonizing}
\bibfield{author}{\bibinfo{person}{Vivetha Thambinathan} {and}
  \bibinfo{person}{Elizabeth~Anne Kinsella}.} \bibinfo{year}{2021}\natexlab{}.
\newblock \showarticletitle{Decolonizing methodologies in qualitative research:
  Creating spaces for transformative praxis}.
\newblock \bibinfo{journal}{\emph{International Journal of Qualitative
  Methods}}  \bibinfo{volume}{20} (\bibinfo{year}{2021}),
  \bibinfo{pages}{16094069211014766}.
\newblock


\bibitem[Tripathi et~al\mbox{.}(2022)]%
        {P27}
\bibfield{author}{\bibinfo{person}{Sandhya Tripathi},
  \bibinfo{person}{Bradley~A. Fritz}, \bibinfo{person}{Michael~S. Avidan},
  \bibinfo{person}{Yixin Chen}, {and} \bibinfo{person}{Christopher~R. King}.}
  \bibinfo{year}{2022}\natexlab{}.
\newblock \showarticletitle{Algorithmic Bias in Machine Learning Based Delirium
  Prediction}.
\newblock \bibinfo{journal}{\emph{ArXiv}}  \bibinfo{volume}{abs/2211.04442}
  (\bibinfo{year}{2022}).
\newblock


\bibitem[Wang et~al\mbox{.}(2022a)]%
        {wang2022TI}
\bibfield{author}{\bibinfo{person}{Angelina Wang}, \bibinfo{person}{Vikram~V
  Ramaswamy}, {and} \bibinfo{person}{Olga Russakovsky}.}
  \bibinfo{year}{2022}\natexlab{a}.
\newblock \showarticletitle{Towards Intersectionality in Machine Learning:
  Including More Identities, Handling Underrepresentation, and Performing
  Evaluation}. In \bibinfo{booktitle}{\emph{2022 ACM Conference on Fairness,
  Accountability, and Transparency}} (Seoul, Republic of Korea)
  \emph{(\bibinfo{series}{FAccT '22})}. \bibinfo{publisher}{Association for
  Computing Machinery}, \bibinfo{address}{New York, NY, USA},
  \bibinfo{pages}{336–349}.
\newblock
\showISBNx{9781450393522}
\urldef\tempurl%
\url{https://doi.org/10.1145/3531146.3533101}
\showDOI{\tempurl}


\bibitem[Wang et~al\mbox{.}(2022b)]%
        {P1}
\bibfield{author}{\bibinfo{person}{Angelina Wang}, \bibinfo{person}{Vikram~V.
  Ramaswamy}, {and} \bibinfo{person}{Olga Russakovsky}.}
  \bibinfo{year}{2022}\natexlab{b}.
\newblock \showarticletitle{Towards Intersectionality in Machine Learning:
  Including More Identities, Handling Underrepresentation, and Performing
  Evaluation}.
\newblock \bibinfo{journal}{\emph{2022 ACM Conference on Fairness,
  Accountability, and Transparency}} (\bibinfo{year}{2022}).
\newblock


\bibitem[Waseem et~al\mbox{.}(2021)]%
        {Waseem2021DisembodiedML}
\bibfield{author}{\bibinfo{person}{Zeerak Waseem}, \bibinfo{person}{Joachim
  Bingel}, {and} \bibinfo{person}{Isabelle Augenstein}.}
  \bibinfo{year}{2021}\natexlab{}.
\newblock \showarticletitle{Disembodied Machine Learning: On the Illusion of
  Objectivity in NLP}.
\newblock \bibinfo{journal}{\emph{ArXiv}}  \bibinfo{volume}{abs/2101.11974}
  (\bibinfo{year}{2021}).
\newblock


\bibitem[Yang et~al\mbox{.}(2020a)]%
        {P25}
\bibfield{author}{\bibinfo{person}{Ke Yang}, \bibinfo{person}{Biao Huang},
  \bibinfo{person}{Julia Stoyanovich}, {and} \bibinfo{person}{Sebastian
  Schelter}.} \bibinfo{year}{2020}\natexlab{a}.
\newblock \showarticletitle{Fairness-Aware Instrumentation of
  Preprocessing~Pipelines for Machine Learning}.
\newblock


\bibitem[Yang et~al\mbox{.}(2020b)]%
        {P12}
\bibfield{author}{\bibinfo{person}{Ke Yang}, \bibinfo{person}{Joshua~R.
  Loftus}, {and} \bibinfo{person}{Julia Stoyanovich}.}
  \bibinfo{year}{2020}\natexlab{b}.
\newblock \showarticletitle{Causal intersectionality for fair ranking}.
\newblock \bibinfo{journal}{\emph{ArXiv}}  \bibinfo{volume}{abs/2006.08688}
  (\bibinfo{year}{2020}).
\newblock


\end{thebibliography}

\clearpage
\appendix
\section*{Appendix}

\section{Tagging for Intersectionality Literature}
\label{app: tagging}
Initially, we only tagged works as incorporating intersectionality literature when they included \citet{collins2020intersectionality}, \citet{crenshaw1989demarginalizing}, \citet{Hancock_2007}, or \citet{Cho_Crenshaw_McCall_2013}. However, during our weekly discussions, we noticed that papers cited a wider array of intersectionality works; either other works by these same authors, or other scholars who center intersectionality within critical disciplines. Because we want to gauge how AI fairness conceptualizes intersectionality, casting a wider net on tags is valuable in that we can include works that, while unaware of our initial list of texts, state intersectionality as a motivation in their work and cite other works about intersectionality like \cite{Collins2017BlackFT}, \cite{Collins2019IntersectionalityAC}, \cite{Collins2015IntersectionalitysDD}, or \cite{Crenshaw_1991}. As a result, we tag the presence of intersectionality literature if any paper includes works that: 1) discuss intersectionality outside of CS and 2) frames intersectionality as a critical social inquiry and praxis framework.

\section{Guiding Questions and Considerations}
\label{app: questions}

\begin{table*}[!ht]
\centering
    \small
    \caption{\citeauthor{collins2020intersectionality}'s tenets of intersectionality and our corresponding guiding questions}
    \vspace{-0.25cm}  
    \begin{tabular}{p{0.15\linewidth}|p{0.75\linewidth}}
    \toprule
    \textbf{Tenet} & \textbf{Guiding Questions} \\
    \midrule
        Social inequality & 1) Do the authors ground their work in how specific social or historical contexts factor into social inequality? \\ 
        & 2) Do the authors acknowledge the implications of their work with respect to social inequality? \\ 
        & 3) Is there a discussion of how intersecting power relations produce social inequality? \\ 
        \midrule
        Social power & 1) Do the authors mention power? \\ 
        & 2) Do the authors discuss any movement of power to the powerless? \\ 
        & 3) Do the authors mention the mutual construction of power? \\ 
        & 4) Is their own power in the work named or do the authors reflexively comment on the oppressive power relations within which their work participates? \\ 
        \midrule
        Social Context & 1) Do the authors name their social context or social location with respect to their work? \\ 
        & 2) Do the authors discuss how their social context influences their ideas and work’s design, decisions, and development? \\ 
        & 3) Do they acknowledge the limitations of their contexts? \\ 
        \midrule
        Relationality & 1) Do the authors discuss the relationships between either social groups or structures? \\ 
        & 2) Do the authors engage with how different social groups, typically treated as separate, face shared oppression? \\ 
        & 3) Do the authors comment on how their identities shape their inquiry in relation to the people affected by their work? \\ 
        \midrule
        Complexity & 1) Do the authors consider cross-sectional social categories? \\ 
        & 2) Do they involve those without power in the generation and social construction of new knowledge? \\ 
        & 3) Do the authors comment on the interplay between technical interventions and social action, or critical inquiry and practice? \\ 
        & 4) Are there any discussions on how spaces operate at different domains of power ? \\ 
        \midrule
        Social justice & 1) Do the authors state their commitment or motivation as social justice? \\ 
        & 2) Do the authors discuss ways in which fair predictions or rules are not equally applied to everyone and can still produce unfair and unequal outcomes? \\ 
        & 3) Do authors aim to dismantle a form of injustice, rather than solely documenting it in the form of a paper? \\ 
        \bottomrule
    \end{tabular}
    \label{tbl:guiding_questions}
\end{table*}

We chose to create 3-4 guiding questions per tenet in order to balance in-depth coverage of each tenet with annotation feasibility. We share all our guiding questions in Table \ref{tbl:guiding_questions}. While some guiding questions are straightforward (e.g., ``Do the authors consider cross-sectional social categories?''), others are more up to our interpretation and experiences (e.g., ``Are there any discussions on how spaces operate at different domains of power?''). Our interpretation of the intersectionality tenets for advancing justice in AI fairness is influenced by our social context and location, including our formal AI training, social identities, and experienced social inequalities (\S\ref{sec:positionality}). For instance, we (the investigators) are all trans and people of color, and hence were likely more attuned to the discussion in \citet{P3} of power differentials in ``regular'' experiences, e.g., going through airport security.

\section{Annotation Methodology}
\label{app:annot_meth}

We follow Lincoln and Guba's 1981 model of trustworthiness in our analysis \cite{nowell2017thematic}, taking steps to maximize its credibility, dependability, confirmability, and transferability.

\begin{itemize}
    \item \textbf{Credibility:} We are highly familiar with \citeauthor{collins2020intersectionality}'s tenets. We also engaged in in-depth intersectionality intensives hosted by Black feminist scholars. Furthermore, we all have done justice work in some capacity. The majority of authors on this paper are trans people of color operating in AI. One author is a social scientist who confronts social inequities in their scholarship through  intersectional perspectives. We spent over 6 months developing the guiding questions. 
    
    \item \textbf{Dependability:} 11 out of the 30 papers were evaluated by three annotators, and we present our tenet-level interannotater agreement for these papers in Table~\ref{tbl:inter_annot}. The scores in Table~\ref{tbl:inter_annot} indicate moderate to high interannotator agreement. The remaining 19 papers were each evaluated by at least 1 annotator.
    \item \textbf{Confirmability:} During weekly investigator meetings, we discussed our guiding questions and identified major sources of disagreement in our annotations.
    \item \textbf{Transferability:} Our guiding questions can be operationalized across paper types and domains outside of academia.
\end{itemize}

\section{Measuring Interannotator Agreement}
We use Randolph's $\kappa$ to estimate inter-annotator agreement, which is free-marginal rather than fixed-marginal; we choose this because $\kappa$ is computed over six distinct items (i.e., tenets).

\begin{table*}[!ht]
\centering
\small
\caption{Tenet-level interannotator scores by Randolph's $\kappa$ and \% agreement}
\vspace{-0.25cm}  

\begin{tabular}{p{0.48\linewidth} p{0.20\linewidth} p{0.2\linewidth}}
\toprule
Paper &  $\kappa$ &  \% agreement \\
\midrule
\citet{P1}        &    1.0000 &    100.00 \\
\citet{P2}        &    1.0000 &    100.00 \\
\citet{P3}        &    0.7778 &     83.33 \\
\citet{P5}        &    1.0000 &    100.00 \\
\citet{P6}        &    0.5556 &     66.67 \\
\citet{P9}        &    0.5556 &     66.67 \\
\citet{P10}       &    0.5556 &     66.67 \\
\citet{P13}       &    1.0000 &    100.00 \\
\citet{P15}       &    0.5556 &     66.67 \\
\citet{P19}       &    0.7778 &     83.33 \\
\citet{P26}       &    0.7778 &     83.33 \\
\bottomrule
\textbf{Average:} &    0.7778 &     83.33 \\
\bottomrule
\end{tabular}
\label{tbl:inter_annot}
\end{table*}

\begin{table*}[!ht]
    \centering
    \caption{Papers with AI fairness research methodology tags}
    \label{tbl:meth_tags}
    \resizebox{.8\linewidth}{!}{
    \begin{tabular}{p{0.05\linewidth}|p{0.15\linewidth}|p{0.1\linewidth}|p{0.2\linewidth}|p{0.15\linewidth}|p{0.1\linewidth}}
    \toprule
        \textbf{ID} & \textbf{Paper} & \textbf{Source of Bias} & \textbf{Intersectonality Operationalization} & \textbf{CS Paper Type} & \textbf{Synergy} \\ 
    \midrule
        1 & \citet{P1} & statistical & full pipeline & empirical & yes \\ 
    \midrule
        2 & \citet{P2} & statistical & in-processing & theoretical, engineering, empirical & no \\ 
    \midrule
        3 & \citet{P3} & systemic & processes & other & yes \\ 
    \midrule 
        4 & \citet{P4} & both & post-processing & empirical & no \\ 
    \midrule
        5 & \citet{P5} & both & in-processing & theoretical, engineering, empirical & yes \\ 
    \midrule
        6 & \citet{P6} & systemic & post-processing & empirical & yes \\ 
    \midrule
        7 & \citet{P7} & systemic & processes, full pipeline & empirical & yes \\ 
    \midrule
        8 & \citet{P8} & systemic & processes & empirical, other & yes \\
    \midrule
        9 & \citet{P9} & statistical & full pipeline & engineering, empirical & no \\ 
    \midrule
        10 & \citet{P10} & both & post-processing & empirical & no \\ 
    \midrule
        11 & \citet{P11} & both & post-processing & empirical & no \\ 
    \midrule
        12 & \citet{P12} & both & full pipeline & theoretical, engineering, empirical & no \\ 
    \midrule
        13 & \citet{P13} & statistical & pre-processing, processes & engineering, empirical & yes \\ 
    \midrule
        14 & \citet{P14} & statistical & in-processing & theoretical, engineering, empirical & no \\ 
    \midrule
        15 & \citet{P15} & systemic & post-processing, processes & theoretical,  empirical & yes \\ 
    \midrule
        16 & \citet{P16} & systemic & processes & theoretical & yes \\ 
    \midrule
        17 & \citet{P17} & both & post-processing & empirical & yes \\
    \midrule
        18 & \citet{P18} & systemic & processes & other & yes \\ 
    \midrule
        19 & \citet{P19} & systemic & post-processing, processes & theoretical, empirical & yes \\ 
    \midrule
        20 & \citet{P20} & statistical & post-processing & engineering, empirical & no \\ 
    \midrule
        21 & \citet{P21} & statistical & in-processing & theoretical,  engineering, empirical & no \\ 
    \midrule
        22 & \citet{P22} & statistical & pre-processing & engineering & no \\
    \midrule
        23 & \citet{P23} & both & processes & other & yes \\ 
    \midrule
        24 & \citet{P24} & both & post-processing & empirical & yes \\
    \midrule
        25 & \citet{P25} & statistical & full pipeline & engineering, empirical & no \\ 
    \midrule
        26 & \citet{P26} & statistical & pre-processing & theoretical, engineering, empirical & no \\ 
    \midrule
        27 & \citet{P27} & both & pre-processing & empirical & yes \\ 
    \midrule
        28 & \citet{P28} & systemic & pre-processing & theoretical, empirical & no \\ 
    \midrule
        29 & \citet{P29} & systemic & processes & other & yes \\ 
    \midrule
        30 & \citet{P30} & systemic & post-processing & other & no \\ 
    \bottomrule
    \end{tabular}
}
\end{table*}

\begin{table*}[!ht]
    \centering
     \caption{Papers with intersectionality-related reference tags}
    \label{tbl:inter_tags}
    \resizebox{.8\linewidth}{!}{
    \begin{tabular}{p{0.05\linewidth} p{0.25\linewidth} p{0.17\linewidth} p{0.15\linewidth} p{0.2\linewidth}}
    \toprule
        ID & Paper & Cites Intersectionality Literature & Says ``Intersectional'' & Says ``Intersectionality'' \\ 
    \midrule
        1 & \citet{P1} & Yes & Yes & Yes \\ 
        2 & \citet{P2} & Yes & Yes & Yes \\ 
        3 & \citet{P3} & Yes & Yes & Yes \\ 
        4 & \citet{P4} & No & Yes & Yes \\ 
        5 & \citet{P5} & Yes & Yes & Yes \\ 
        6 & \citet{P6} & Yes & Yes & Yes \\ 
        7 & \citet{P7} & Yes & Yes & Yes \\ 
        8 & \citet{P8} & Yes & Yes & Yes \\ 
        9 & \citet{P9} & Yes & Yes & Yes \\ 
        10 & \citet{P10} & Yes & Yes & Yes \\ 
        11 & \citet{P11} & No & Yes & No \\ 
        12 & \citet{P12} & Yes & Yes & Yes \\ 
        13 & \citet{P13} & No & Yes & Yes \\ 
        14 & \citet{P14} & Yes & Yes & Yes \\ 
        15 & \citet{P15} & Yes & Yes & Yes \\ 
        16 & \citet{P16} & Yes & Yes & Yes \\ 
        17 & \citet{P17} & Yes & Yes & Yes \\ 
        18 & \citet{P18} & Yes & Yes & Yes \\ 
        19 & \citet{P19} & Yes & Yes & No \\ 
        20 & \citet{P20} & No & Yes & Yes \\ 
        21 & \citet{P21} & No & Yes & No \\ 
        22 & \citet{P22} & No & Yes & No \\ 
        23 & \citet{P23} & Yes & No & Yes \\ 
        24 & \citet{P24} & Yes & Yes & Yes \\ 
        25 & \citet{P25} & Yes & Yes & Yes \\ 
        26 & \citet{P26} & Yes & Yes & Yes \\ 
        27 & \citet{P27} & No & No & Yes \\ 
        28 & \citet{P28} & Yes & Yes & Yes \\ 
        29 & \citet{P29} & No & Yes & Yes \\ 
        30 & \citet{P30} & Yes & No & Yes \\ 
    \end{tabular}%
    }
\end{table*}

\end{document}